\documentclass[manuscript]{aastex}
\usepackage{amsfonts,amsmath,graphicx,listings,hyperref}
\pdfoutput=1

\begin{document}

\title{Pulsation period variations in the RRc Lyrae star KIC 5520878}

\author{Michael Hippke}
\email{hippke@ifda.eu}
\affil{Institute for Data Analysis, Luiter Str. 21b, 47506 Neukirchen-Vluyn, Germany}

\author{John G. Learned}
\email{jgl@phys.hawaii.edu}
\affil{High Energy Physics Group, Department of Physics and Astronomy, University of Hawaii, Manoa 327 Watanabe Hall, 2505 Correa Road Honolulu, Hawaii 96822 USA}

\author{A. Zee}
\email{zee@kitp.ucsb.edu}
\affil{Kavli Institute for Theoretical Physics, University of California, Santa Barbara, California 93106 U.S.A.}

\author{William H. Edmondson}
\email{w.h.edmondson@bham.ac.uk}
\affil{School of Computer Science, University of Birmingham, Birmingham B15 2TT, UK}

\author{John F. Lindner}
\email{jlindner@wooster.edu}
\affil{Physics Department, The College of Wooster, Wooster, Ohio 44691, USA}
\affil{Department of Physics and Astronomy, University of Hawai`i at M\=anoa, Honolulu, Hawai`i 96822, USA}

\author{Behnam Kia}
\email{behnam@hawaii.edu}
\affil{Department of Physics and Astronomy, University of Hawai`i at M\=anoa, Honolulu, Hawai`i 96822, USA}

\author{William L. Ditto}
\email{wditto@hawaii.edu}
\affil{Department of Physics and Astronomy, University of Hawai`i at M\=anoa, Honolulu, Hawai`i 96822, USA}

\author{Ian R. Stevens}
\email{irs@star.sr.bham.ac.uk}
\affil{School of Physics and Astronomy, University of Birmingham, Birmingham B15 2TT, UK}

\begin{abstract}
Learned et al. proposed that a sufficiently advanced extra-terrestrial civilization may tickle Cepheid and RR Lyrae variable stars with a neutrino beam at the right time, thus causing them to trigger early and jogging the otherwise very regular phase of their expansion and contraction. This would turn these stars into beacons to transmit information throughout the galaxy and beyond. The idea is to search for signs of phase modulation (in the regime of short pulse duration) and patterns, which could be indicative of intentional, omnidirectional signaling. 

We have performed such a search among variable stars using photometric data from the Kepler space telescope. In the RRc Lyrae star KIC 5520878, we have found two such regimes of long and short pulse durations. The sequence of period lengths, expressed as time series data, is strongly auto correlated, with correlation coefficients of prime numbers being significantly higher ($p=99.8$\%). Our analysis of this candidate star shows that the prime number oddity originates from two simultaneous pulsation periods and is likely of natural origin. 

Simple physical models elucidate the frequency content and asymmetries of the KIC 5520878 light curve.

Despite this SETI null result, we encourage testing other archival and future time-series photometry for signs of modulated stars. This can be done as a by-product to the standard analysis, and even partly automated. 
\end{abstract}


\section{Introduction}
We started this work with the simple notion to explore the phases of variable stars, given the possibility of intentional modulation by some advanced intelligence \citep{Learned2008}. The essence of the original idea was that variable stars are a natural target of study for any civilization due to the fact of their correlation between period and total light output. This correlation has allowed them to become the first rung in the astronomical distance ladder, a fact noted in 1912 by Henrietta Leavitt \citep{Pickering1912}. Moreover, such oscillators are sure to have a period of instability at which time they are sensitive to perturbations, and which can be triggered to flare earlier in their cycle than otherwise. If some intelligence can do this in a regular manner, they have the basis of a galactic semaphore for communicating not only within one galaxy, but in the instance of brighter Cepheids with many galaxies. This would be a one-way communication, like a radio station broadcasting indiscriminately. One may speculate endlessly about the motivation for such communication; we only aver that there are many possibilities and given our complete ignorance of any other intelligence in the universe, we can only hope we will recognize signs of artificiality when we see them. 

We must be careful to acknowledge that we are fully aware that this is a mission seemingly highly unlikely to succeed, but the payoff could be so great it is exciting to try. In what we report below, we have begun such an investigation, and found some behaviour in a variable star that is, at first glance, very hard to understand as a natural phenomenon. A deeper analysis of simultaneous pulsations then shows its most likely natural origin. This opens the stage for a discussion of what kind of artificiality we should look for in future searches. 

By way of more detailed introduction, astronomers who study variable stars in the past have had generally sparsely sampled data, and in earlier times even finding the periodicity was an accomplishment worth publishing. Various analytical methods were used, such as the venerable 'string method', which involves folding the observed magnitudes to a common one cycle graph, phase ordering and connecting the points by a virtual 'string' \citep{Petrie1962, Dworetzki1982}. Editing of events which did not fit well was common. One moved the trial period until the string length was minimized. In more recent times with better computing ability, the use of Lomb-Scargle folding \citep{Lomb1976, Scargle1982} and Fourier transforms \citep{Ransom2010} has become prominent. For our purposes however, such transforms obliterate the cycle to cycle variations we seek, and so we have started afresh looking at the individual peak-to-peak periods and their modulation. 

The type of analysis we want to carry out could not have been done even a few years ago, since it requires many frequent observations on a single star, with good control over the photometry over time. This is very hard to do from earth -- we seem to be, accidentally, on the wrong planet for the study of one major type of variables, the RR Lyrae stars \citep{Welch2014}. Typically, these stars pulsate with periods near 0.5 days (a few have shorter or longer periods). This is unfortunately
well-matched to the length of a night on earth. Even in case of good weather, and when observing every night from the same location, still every second cycle is lost due to it occurring while the observation site is in daylight. This changed with the latest generation of space telescopes, such as \textit{Kepler} and \textit{CoRoT}. The \textit{Kepler} satellite was launched in 2009 with a primary mission of detecting the photometric signatures of planets transiting stars. \textit{Kepler} had a duty cycle of up to 99\%, and was pointed at the same sky location at all times. The high-quality, near uninterrupted data have also been used for other purposes, such as studying variable stars. One such finding was a remarkable, but previously completely undetected and unsuspected behavior: period-doubling. This is a two-cycle modulation of the light curve of Blazhko-type RR Lyrae stars. And it is not a small effect: In many cases, every other peak brightness is different from the previous one by up to 10\% \citep{Blazhko1907, Smith2004, Kolenberg2004, Szabo2010, Kollath2011, Smolec2012}. This has escaped earth-bound astronomers for reasons explained above\footnote{It has also led to a campaign of observing RR Lyrae, the prototype star, from different Earth locations, in order to increase the coverage \citep{Borgne2014}. This is especially useful as \textit{Kepler} finished its regular mission due to a technical failure.}.

\section{Data set}
Our idea was to search for period variations in variable stars. We employed the database best suited for this today, nearly uninterrupted photometry from the \textit{Kepler} spacecraft, covering 4 years of observations with ultra-high precision \citep{Caldwell2010}. In what follows, we describe this process using RRc Lyrae KIC 5520878
as an example. For this star, we have found the most peculiar effects which are described in this paper. For comparison, we have also applied the same analysis for other variable stars, as will be described in the following sections.

KIC 5520878 is a RRc Lyrae star with a period of $P_{1}$=0.26917082 days (the number of digits giving a measure of the accuracy of the value \citep{Nemec2013}). It lies in the field of view of the \textit{Kepler} spacecraft, together with 40 other RR Lyrae stars, out of which 4 are of subtype RRc\footnote{These are KIC 4064484, 5520878, 8832417, and 9453114, according to \citet{Nemec2013}.}. Only one Cepheid is known in the area covered by \textit{Kepler} \citep{Szabo2011}. The brightness of KIC5520878, measured as the \textit{Kepler magnitude}, is \textit{Kp}=14.214.

\subsection{Data selection and retrieval}
The \textit{Kepler} data were downloaded from the Mikulski Archive for Space Telescopes (MAST) in FITS format and all files were merged with \textit{TopCat} (Bristol University). The data come with adjustments for barycentering (and the corrections applied), and flux data are available as raw simple aperture photometry (called \textit{SAP} by the Kepler team). There is also a pre-cleaned version available called \textit{PDCSAP}, which includes corrections for drift, discontinuities and events such as cosmic rays. The Kepler Data Characteristics Handbook \citep{Handbook2011} points out that this cleaning process ``[does not] perfectly preserve general stellar variability'' and investigators should use their own cleaning routines where needed.

For the period length analysis explained in the next sections, we have tried both \textit{SAP} and \textit{PDCSAP} data, as well as applied our own correction obtained from nearby reference stars. All three data sets gave identical results. This is due to the corrections being much smaller than intrinsic stellar variations: While instrumental amplitude drifts are in the range of a few percent over the course of months, the star's amplitude changes by $\sim$40\% during its 0.269d (6.5hr) cycle. For a detailed study of best-practice detrending in RR Lyrae stars, we refer to \citet{Benko2014a}.

\subsection{Data processing: Measuring pulsation period lengths}
Most commonly, Fourier decomposition is used to determine periodicity. Other methods are phase dispersion minimization (PDM), fitting polynomials, or smoothing for peak/low detection. The Fourier transformation (FT) and its relatives (such as the least-squares spectral analysis used by Lomb--Scargle for reduction of noise caused by large gaps) have their strength in showing the total spectrum. FT, however, cannot deliver clear details about trends in or evolution
of periodicity. This is also true for PDM, as it does not identify pure phase variations, since it uses the whole light curve to obtain the solution \citep{Stellingwerf2013}. 

The focus of our study was pulsation period length, and its trends. We have therefore tried different methods to detect individual peaks (and lows for comparison). 

In the short cadence data (1 minute integrations), we found that many methods work equally well. As the average period length of KIC 5520878 is $\sim$0.269 days, we have $\sim$387 data points in one cycle. First of all, we tried eye-balling the time of the peaks and noted approximate times. As there is some jitter on the top of most peaks, we then employed a centered moving-average (trying different lengths) to smooth this out, and calculated the maxima of this smoothed curve. Afterwards, we fitted n-th order polynomials, trying different numbers of terms. While this is computationally more expensive, it virtually gave the same results as eye-balling and smoothing. Deviations between the methods are on the order of a few data points (minutes). We judge these approaches to be robust and equally suitable. Finally, we opted to proceed with the peak times derived with the smoothing approach as this is least susceptible for errors. 

\begin{figure}[ht]
\begin{center}
\includegraphics[width=0.8\textwidth]{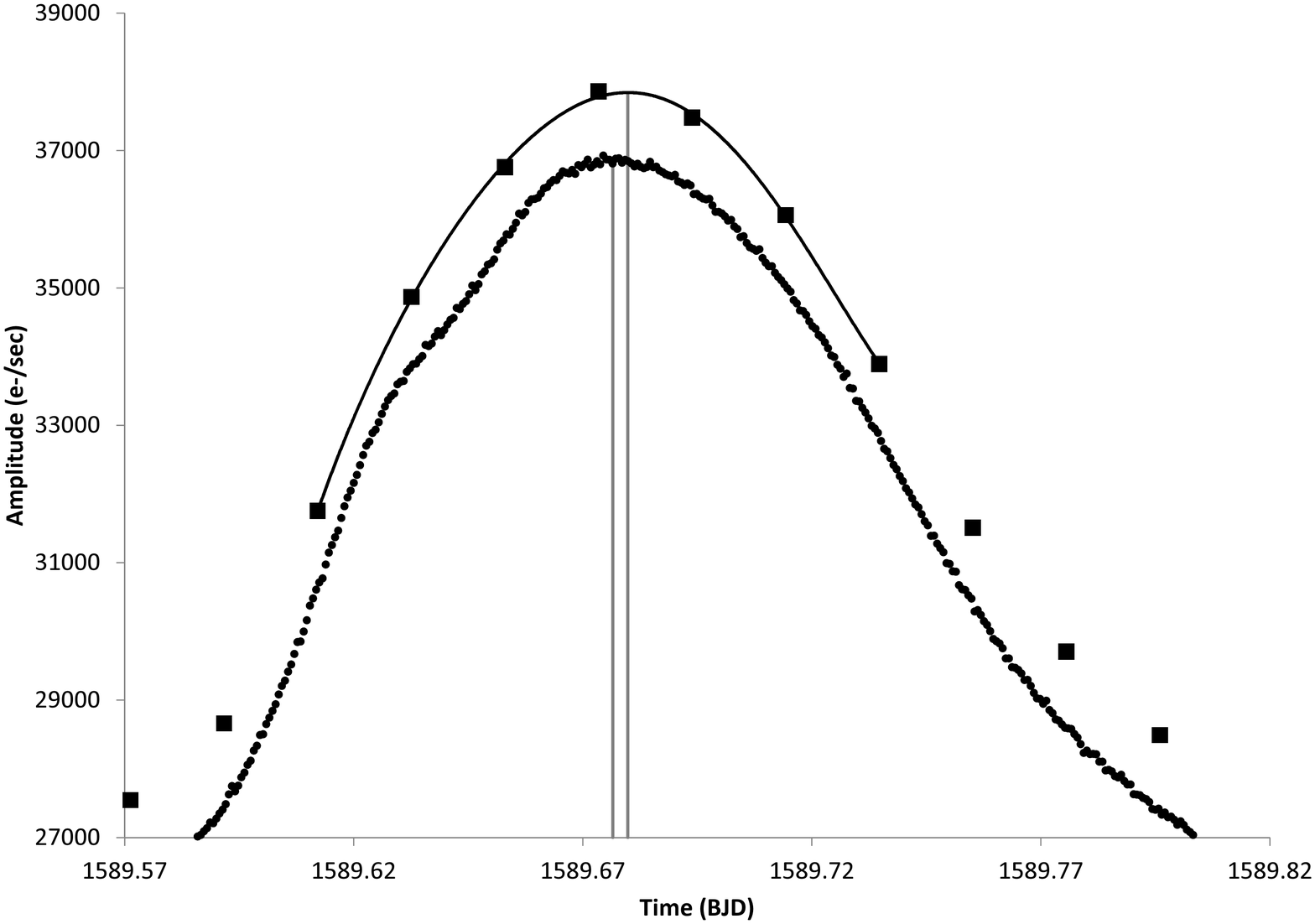}
\end{center}
\caption{\label{fig:f1}Both peak finding methods applied: Square symbols show LC data points, with a $5^{th}$-order polynomial fitted to the 7 highest values. The right of the vertical lines indicates the LC peak from this method. The small dots below are the SC data points, shifted by $1,000 e^{-}$ downwards for better visibility. The peak of the SC data, derived with the smoothing method, is shown by the left of the vertical lines. The two times indicated by the two lines differ by 4 minutes. This is also the average deviation for SC versus LC data. Times are in Barycentric Julian Date (MJD-2454833)}
\end{figure}

In the long cadence data (30 minutes), there are only $\sim$13 data points in one cycle (Fig.~\ref{fig:f1}). Simply using the bin with the highest luminosity
thus introduces large timing errors. We tried fitting templates to the curve, but found this very difficult due to considerable change in the shape of the light curve from cycle to cycle. This stems from two main effects: Amplitude variations of $\sim$3\% cycle-to-cycle, and the seemingly random occurrence of a bump during luminosity increase, when luminosity rises steeply and then takes a short break before reaching its maximum. We found fitting polynomials to be more efficient, and tested the results for varying number of terms and number of data points involved. The benchmark we employed was the short cadence data, which gave us 1,810 cycles for comparison
with the long cadence data. We found the best result to be a 5th-order-polynomial fitted to the 7 highest data points of each cycle, with new parameters for each cycle. This gave an average deviation of only 4.2 minutes, when compared to the short cadence data. Fig.~\ref{fig:f1} shows a typical fit for short and long cadence data. 

\subsection{Data processing result}
During the 4 years of Kepler data, 5,417 pulsation periods passed. We obtained the lengths of 4,707 of these periods (87\%), while 707 cycles (13\%) were lost due to spacecraft downtimes and data glitches.
Out of the 4,707 periods we obtained, we had short cadence data for 1,810 periods (38\%). For the other 62\% of pulsations, we used long cadence data.

\subsection{Data quality}
\begin{figure}[ht]
\begin{center}
\includegraphics[width=\textwidth]{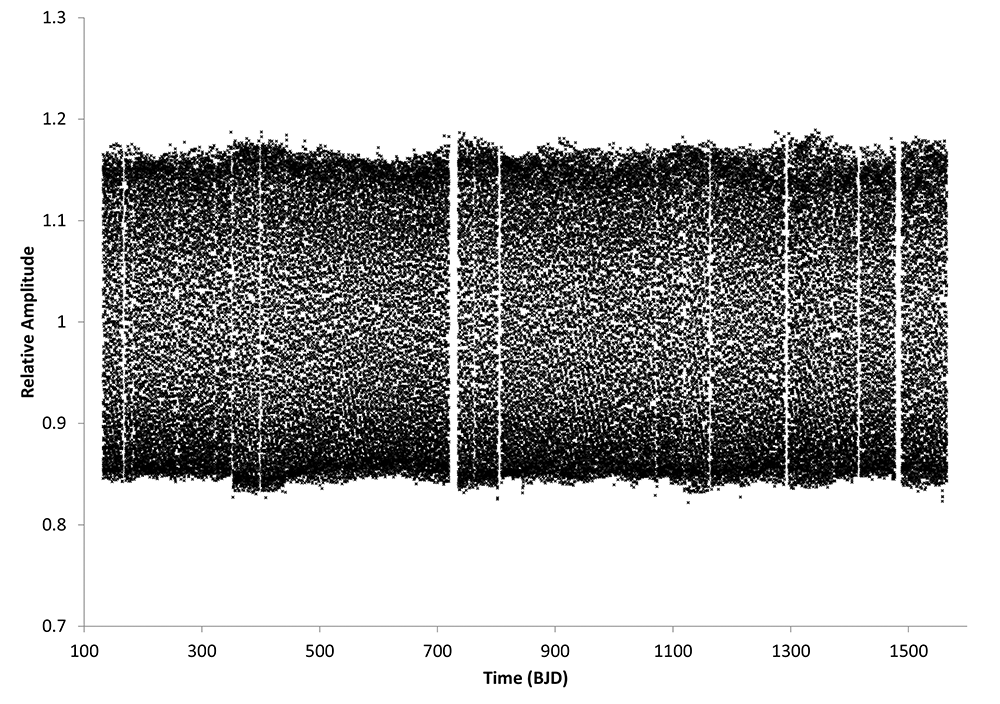}
\end{center}
\caption{\label{fig:f2}Amplitude over time for the complete data set.}
\end{figure}

The errors given by the Kepler team for short integrations (1 minute) for KIC 5520878 are $\leq$35e\textsuperscript{-}/sec for the flux (that is 0.1\% of an average flux of 35,000e\textsuperscript{-}/sec). These errors are smaller than the point size of Fig.~\ref{fig:f2}, which shows the complete data set. We have applied corrections for instrumental drift by matching to nearby stable stars. Some instrumental residuals on the level of a few percent remain, which are very hard to clean out in such a strongly variable star. As explained above, these variations are irrelevant for our period length determination.

\section{Basic facts on RRc Lyrae KIC 5520878}

\subsection{Metallicity and radial velocity}
KIC 5520878 has been found to be metal-rich. When expressed as [Fe/H], which gives the logarithm of the ratio of a star's iron abundance compared to that of our Sun, KIC 5520878 has a ratio of [Fe/H]=-0.36$\pm$0.06, that is about half of that of our sun. This information comes from a spectroscopic study with the Keck-I 10m HIRES echelle spectrograph for the 41 RR Lyrae in the Kepler field of view \citep{Nemec2013}. The study also derived temperatures (KIC 5520878: 7,250 K, the second hottest in the sample) and radial velocity of -0.70$\pm$0.29 km/s.

When compared to the other RR Lyrae from Kepler, or more generally in our solar neighborhood, it becomes clear that metal-rich RR Lyrae are rare. Most (95\%) RR Lyrae seem to be old ($>$10Gyr \citep{Lee1992}), metal-poor stars with high radial velocity \citep{Layden1995}. They belong to the halo population of the galaxy, orbiting around the galactic center with average velocities of $\sim$230km/s. 

\begin{figure}[ht]
\includegraphics[width=\textwidth]{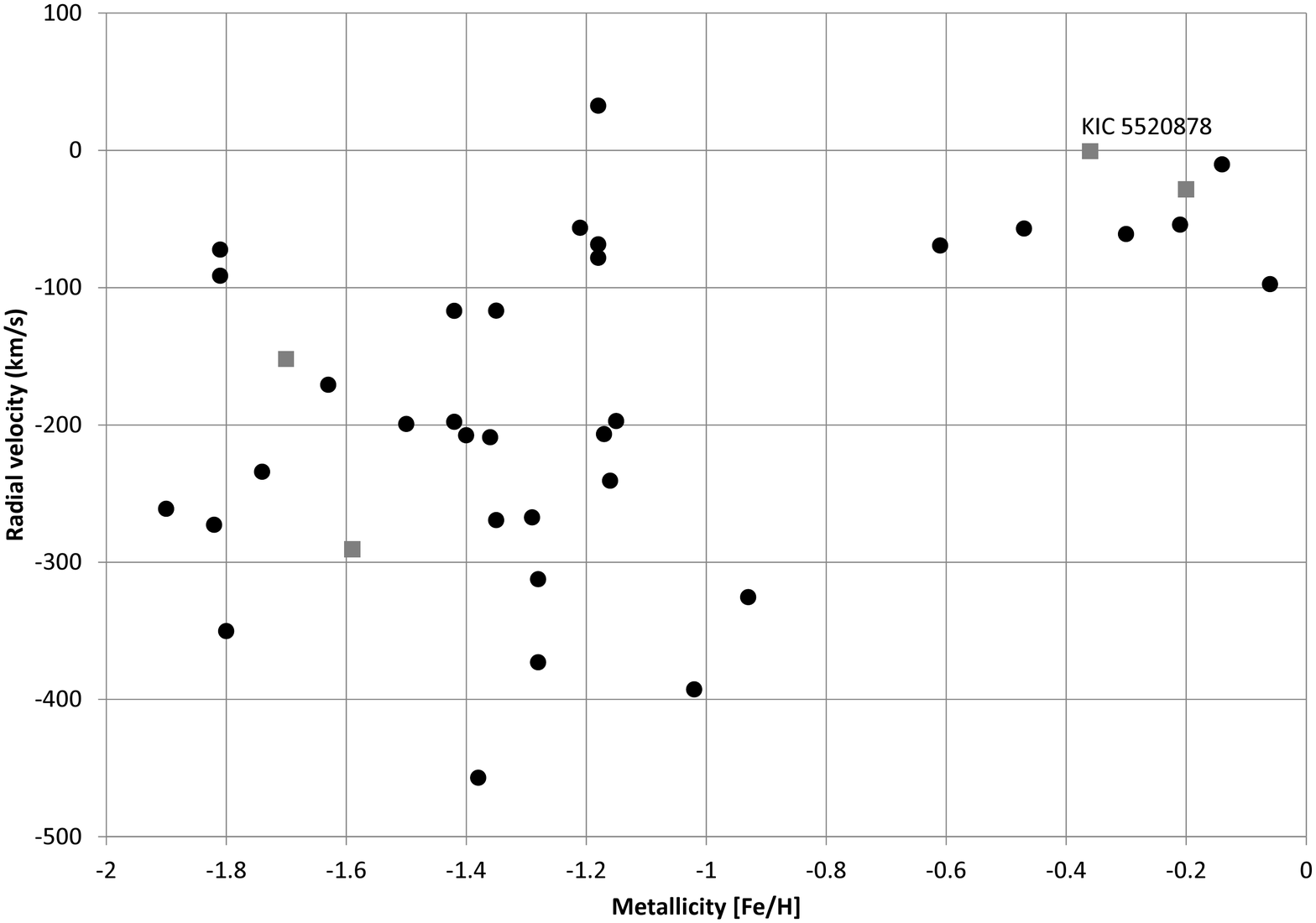}
\caption{\label{fig:f3}Radial velocity versus metallicity for the 41 known Kepler RR Lyrae. Black dots represent sub-type RRab, while gray squares represent sub-type RRc.}
\end{figure}

There seems to exist, however, a second population of RR Lyrae. Their high metallicity is a strong indicator of their younger age, as heavy elements were not common in the early universe. Small radial velocities point towards them co-moving with the galactic rotation and, thus, them belonging to the younger population of the galactic disk, sometimes divided into the very young thin disk and the somewhat older thick disk. 
Our candidate star, RRc KIC 5520878, seems to be one of these rarer objects, as shown in Fig.~\ref{fig:f3}.

\subsection{Cosmic distance}
RR Lyrae stars are part of our cosmic distance ladder, due to the peculiar fact that their absolute luminosity is simply related to their period, and hence if one measures the period one gets the light output. By observing the apparent brightness, one can then derive the distance. The empirical relation is $5log_{10}D=m-M-10$ where $m$ is the apparent magnitude and $M$ the absolute magnitude. If we take \textit{Kp}=14.214 for KIC5520878 and $M=0.75$ for RR Lyrae stars \citep{Layden1995}, the distance calculates to $D=4.9$kpc (or $\sim$16,000 LY), that is $\sim$16\% of the diameter of our galaxy -- the star is not quite in our neighborhood.

\subsection{Rotation period}
Taking an empirical relation of pulsation period $P$ and radius $R$ of RR Lyrae stars \citep{Marconi2005} of $log_{10}⁡R=0.90(\pm0.03)+0.65(\pm0.03)log_{10}⁡P$, one can derive $R=3.4 R_{SUN}$ for KIC 5520878. The inclination angle $i$ is not derivable with the data available. Thus, a corresponding minimum rotation period would be $33.8\pm1$ days at $i=90°$, or 47.8 days at $i=45°$.  The minimum rotation period of 33.8 days is equivalent to $\sim$126 cycles (of 0.269 days length), so that the rotation period is much slower than the pulsation period.

\subsection{Fourier frequencies}
KIC5520878 has been studied in detail \citep{Moskalik2014}. In a Fourier analysis, a secondary frequency $f_{X}=5.879 d^{-1} (P_{X}=0.17$ days) was found. The author calculates a ratio $P_{X}/P_{1} = 0.632$, a rare but known ratio among RR Lyrae stars. This rareness is probably caused by instrumental bias, as these secondary frequencies are low in amplitude, and were only detected with the rise of high-quality time-series photometry (e.g. OGLE) and milli-mag precision space photometry. Indeed, the $\sim$0.6 period ratio was detected in six out of six randomly selected RRc stars observed with space telescopes \citep{Moskalik2014,Szabo2014}.

In addition, significant subharmonics were found with the strongest being $f_{Z} = 2.937 d^{-1} (P_{Z}=0.34$ days), that is at $1/2f_{X}$. Their presence is described as ``a characteristic signature of a period doubling of the secondary mode, $f_{X}$ ($\ldots$). Its origin can be traced to a half-integer resonance between the pulsation modes (Moskalik, Buchler 1990).'' The author also compares several RRc stars and finds ``the richest harvest of low amplitude modes ($\ldots$) in KIC5520878, where 15 such oscillations are detected. Based on the period ratios, one of these modes can be identified with the second radial overtone, but all others must be nonradial. Interestingly, several modes have frequencies significantly lower than the radial fundamental mode. This implies that these nonradial oscillations are not p-modes, but must be of either gravity or mixed mode character.'' These other frequencies are usually said to be non-radial, but could also be explained by ``pure radial pulsation combinations of the frequencies of radial fundamental and overtone modes'' \citep{Benko2014b}. However, this explanation might only apply to fundamental-mode (RRab) stars. Since RRab and RRc stars pulsate in different modes, period ratios in one class might tell little about the period ratios in the other class. The $\sim$0.6 period ratio is close to the fundamental mode-second overtone period ratio, but there are no radial modes that fall to this value against the first overtone, and thus the $P_{X}$-mode is commonly identified as non-radial.

In section~\ref{sec:prime} we explore the Fourier spectrum in more detail.

\subsection{An optical blend?}
\begin{figure}[ht]
\begin{center}
\includegraphics[width=6cm]{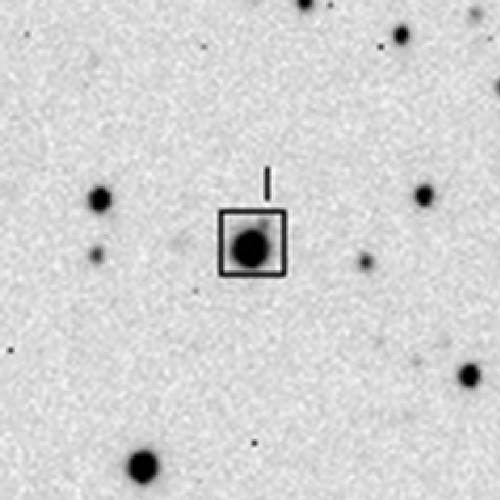}
\end{center}
\caption{\label{fig:f4}CCD photography centered on KIC5520878. The Kepler aperture mask of 4x4 Kepler pixels (1px $\approx3.9$ arcsec) is shown as a square. It fully contains the faint uncataloged star, marked with a vertical line.}
\end{figure}

Unusual frequencies might be contributed by an unresolved companion with a different variability. To check this, we have obtained observations down to magnitude 19.6 with a privately owned telescope and a clear filter. Using a resolution of 0.9px/arcsec, it can be easily seen that a faint (18.6mag) companion is present at a distance of $\approx$7 arcsec, as shown in Fig. ~\ref{fig:f4}. The standard Kepler aperture mask fully contains this companion, however the magnitude difference between this uncataloged companion and KIC5520878 is larger than 4.4 magnitudes, so that the light contribution into our Kepler photometry is very small. At a distance of 4.9kpc, the projected distance of this companion is $\approx$0.15 parsec, resulting in a potential circular orbit of $\approx4\times10^{6}$ years. The most likely case is, however, that this star is simply an unrelated, faint background (or foreground) star with negligible impact to Kepler photometry.

A second argument can be made against blending: With Fourier transforms, linear combinations of frequencies can be found. These are intrinsic behavior of one star's interior, and cannot be produced by two separate stars.

\subsection{Phase fold of the light curve}
\begin{figure}[ht]
\includegraphics[width=\textwidth]{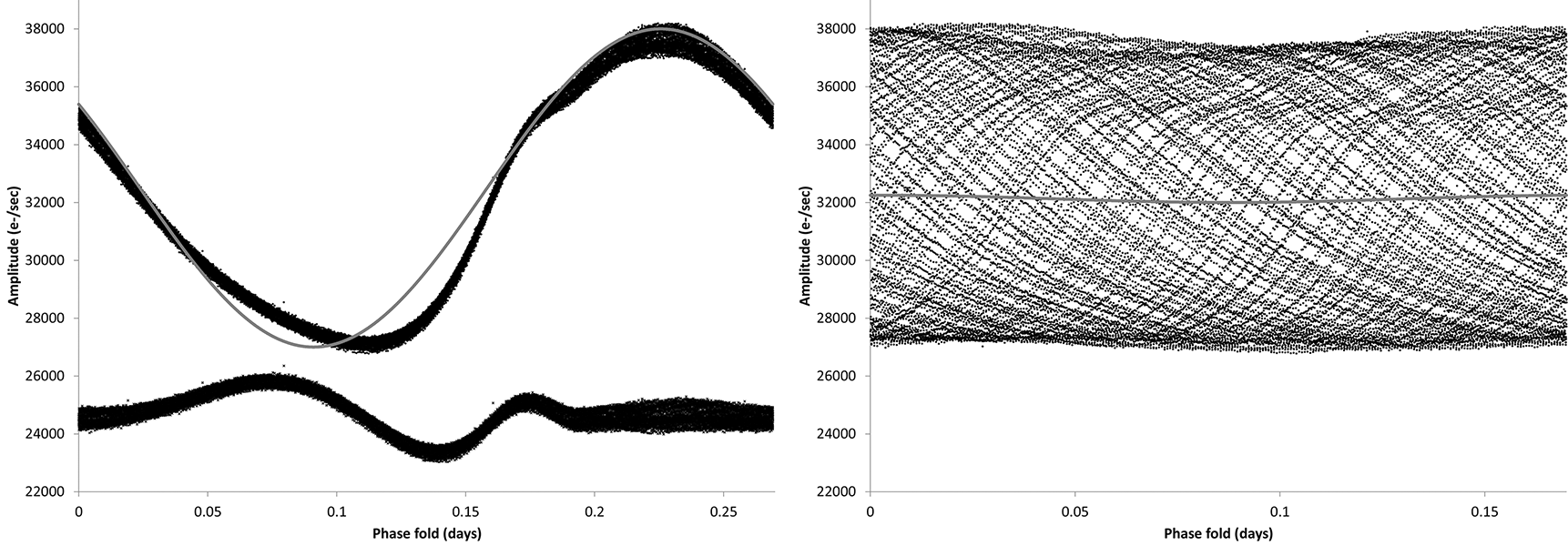}
\caption{\label{fig:f5} Left panel: Folded light curve for main pulsation period $P_{1}$=0.269d with sine fit (grey) centered on peak (upper curves). Below: Residuals of sine fit, displayed at 25,000. Right panel: Phase fold to $P_{2}$=0.17d. Note changed horizontal axis and sine fit (grey) with much smaller variation, about 2\% of $P_{1}$.}
\end{figure}

We have used all short cadence data and created a folded light curve, as shown in Fig.~\ref{fig:f5}. It clearly displays the two periods present: The strong main pulsation $P_{1}$, and the weaker secondary pulsation $P_{X}$.

\subsection{Amplitude features: No period doubling or Blazhko-effect}
\begin{figure}[ht]
\includegraphics[width=\textwidth]{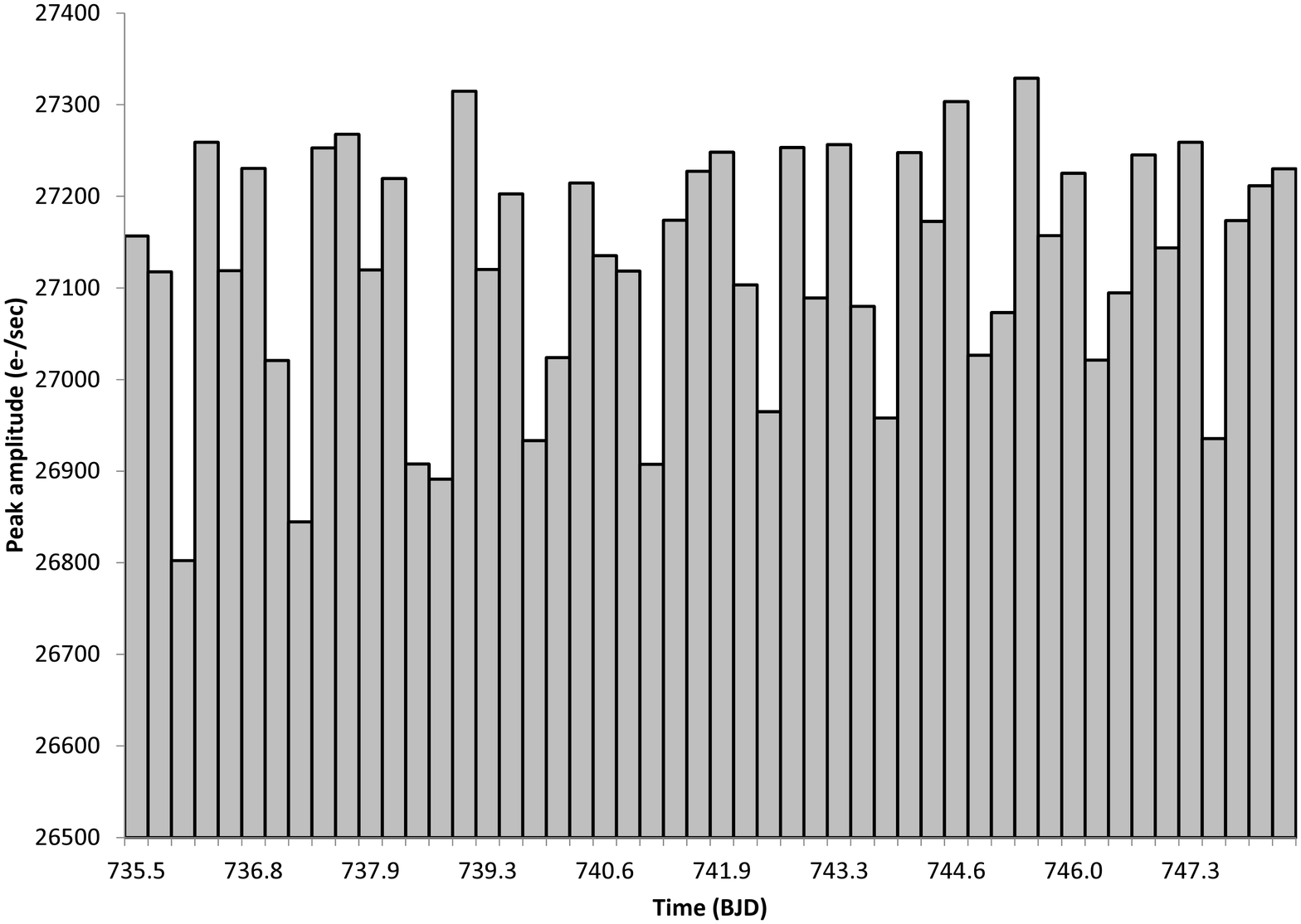}
\caption{\label{fig:f6}Zoom into 50 subsequent amplitudes at peak time. There is no indication of ``period doubling'', the recently discovered cycle-to-cycle variation. The variation present seems to be of a more complex structure.}
\end{figure}

With Kepler data, a previously unknown effect in RR Lyrae was discovered, named ``period doubling'' \citep{Kolenberg2010}. This effect manifests itself in an amplitude variation of up to 10\% of every second cycle. It was found in several Blazhko-RRab stars, but its occurrence rate is yet unclear. Our candidate star, KIC 5520878, does not show the period doubling effect with respect to the main pulsation mode. Judging from Fig.~\ref{fig:f6}, which shows a typical sample of amplitude fluctuations, the variations are more complex. However, as will be explained below, the star does exhibit a strong and complex period variation in the secondary pulsation $P_{X}$.

Regarding the Blazhko effect, a long-term variation best visible in an amplitude-over-time graph, we can see no clear evidence for it being present in KIC 5520878. Fig.~\ref{fig:f2} shows the complete data set, which features some amplitude variations over time, but not of the typical sinusoidal Blazhko style.

\section{Period length and variations}

\subsection{Measured period lengths}
The average period length is constant at $P_{1}$=0.269d when averaged on a timescale of more than a few days. On a shorter timescale, larger variations can be seen. Peak-to-peak, the variation is in the range [0.24d to 0.30d], low-to-low in the range [0.26d to 0.28d]. For any given cycle, the qualitative result of measuring through the lows and through the peaks is equivalent, meaning that a short (long) period is measured to be short (long), no matter whether peaks or lows are used. As the peak-to-peak method gives larger nominal differences (and thus better signal-to-noise), we focus on the peak times from here on.

\begin{figure}[ht]
\begin{center}
\includegraphics[width=0.6\textwidth]{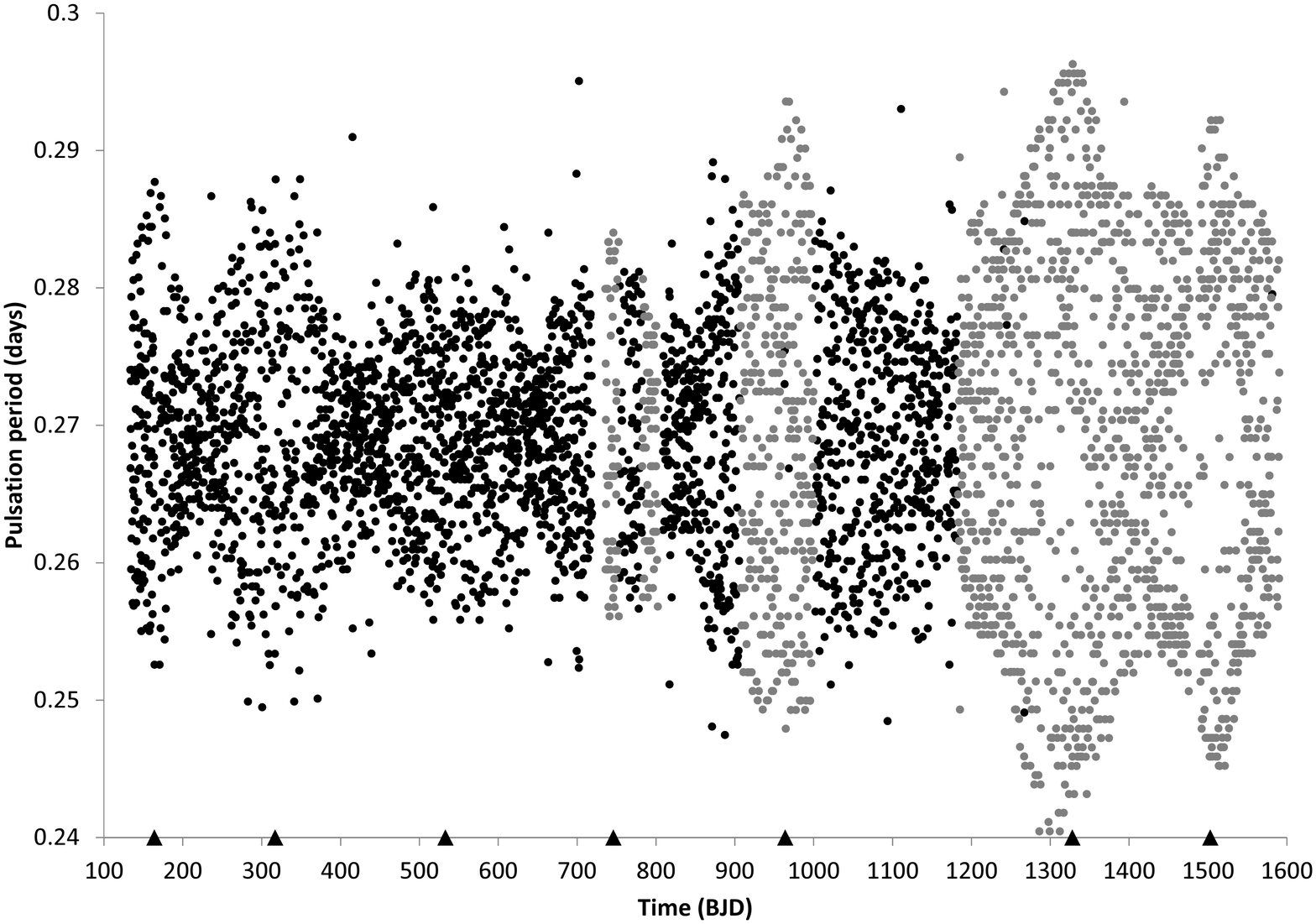}

\includegraphics[width=0.6\textwidth]{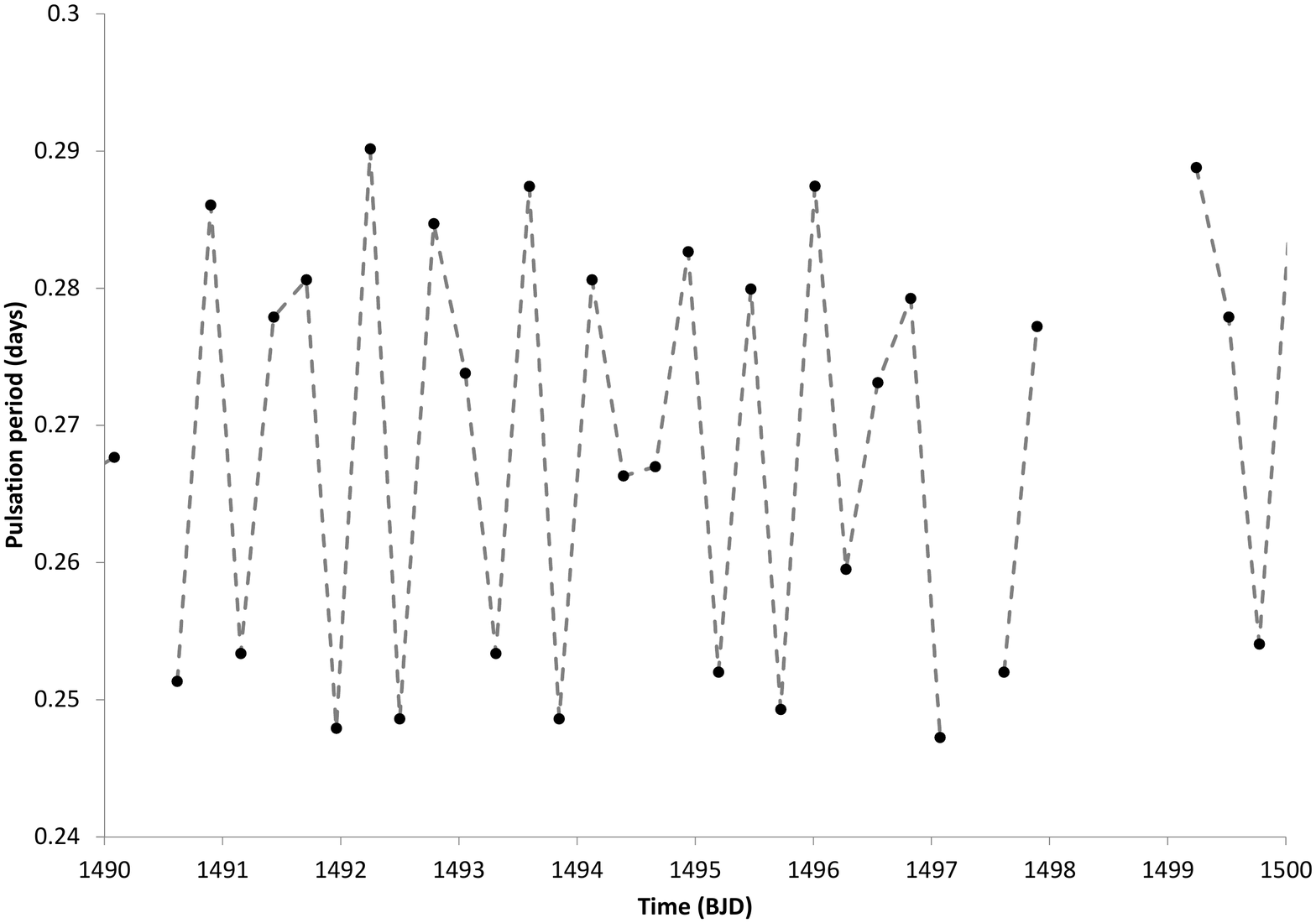}
\end{center}
\caption{\label{fig:f7}Upper panel: Pulsation period lengths over time using LC data (black) and SC data (gray). Note the fluctuations of variance. The peaks are indicated with triangles at $P_{1}$=0.24d. In addition, there are ``forbidden zones'', e.g. at BJD=1500: $P_{1}$=0.265d. This diagram can easily be understood as the typical $O-C$ (observed minus calculated) diagram if the average main pulsation period $P_{1}$=0.269 is set to 0. Periods longer than $P_{1}$ then have positive $O-C$ values and vice versa. Lower panel: Zoom showing the progression of subsequent period lengths.}
\end{figure}

The variation seems to fluctuate in cycles of variable length. We estimate maxima at BJD$\sim$164, 317, 533, 746, 964, 1328 and 1503. Their separation is then (in days): 153, 216, 213, 218, 364 and 175 (Fig.~\ref{fig:f7}). As some of these are near one Earth (Kepler) year, or half of it, we have to carefully judge the data regarding barycentering, which has been performed by the Kepler team. We argue against a barycentering error for three reasons: First, the differences caused by barycentering between two subsequent cycles are negligible ($<0.5$s), while the cycle has an average length of $\sim$387 minutes. Second, the cycles are of varying length, while a year is not. Third, the large Kepler community would most likely already have found such a major error. We therefore judge the long-term variation of period lengths to be a real phenomenon. 

In section~\ref{sec:primeAC}, we will explain how the period length variations originate from the presence of a second pulsation mode.

\subsection{Comparing period length variations to other RR Lyrae stars }
The period variation of RR Lyrae, the prototype star, has been measured at $\sim$0.53\% \citep{Stellingwerf2013}. This is low compared to $\sim$22\% for KIC 5520878. The low value of RR Lyrae is reproducible with our method of individual peak detection. The authors estimate a scatter of $\sim$0.01 days caused by measurement errors. 
We have furthermore looked at several other RR Lyrae stars in the Kepler field of view, and found others (V445=KIC 6186029; KIC 8832417) that also show large period variations. For KIC 6186029 the variations are 0.49d to 0.53d, that is $\sim$8\%. It has been analyzed in a paper whose title summarizes the findings as ``Two Blazhko modulations, a non-radial mode, possible triple mode RR Lyrae pulsation and more'' \citep{Guggenberger2012}.

\subsection{Period-Amplitude relation}
KIC 5520878 shows a significant ($p>99.9$\%) relation of longer periods having higher flux. In a recent paper, a counterclockwise looping evolution between amplitude and period in the prototype RR Lyrae was detected. This looping, so far only found in RRab-Blazhko-stars, cannot be reproduced in KIC 5520878 \citep{Stellingwerf2013}.

\subsection{Distribution of period lengths}
\begin{figure}[ht]
\begin{center}
\includegraphics[width=0.8\textwidth]{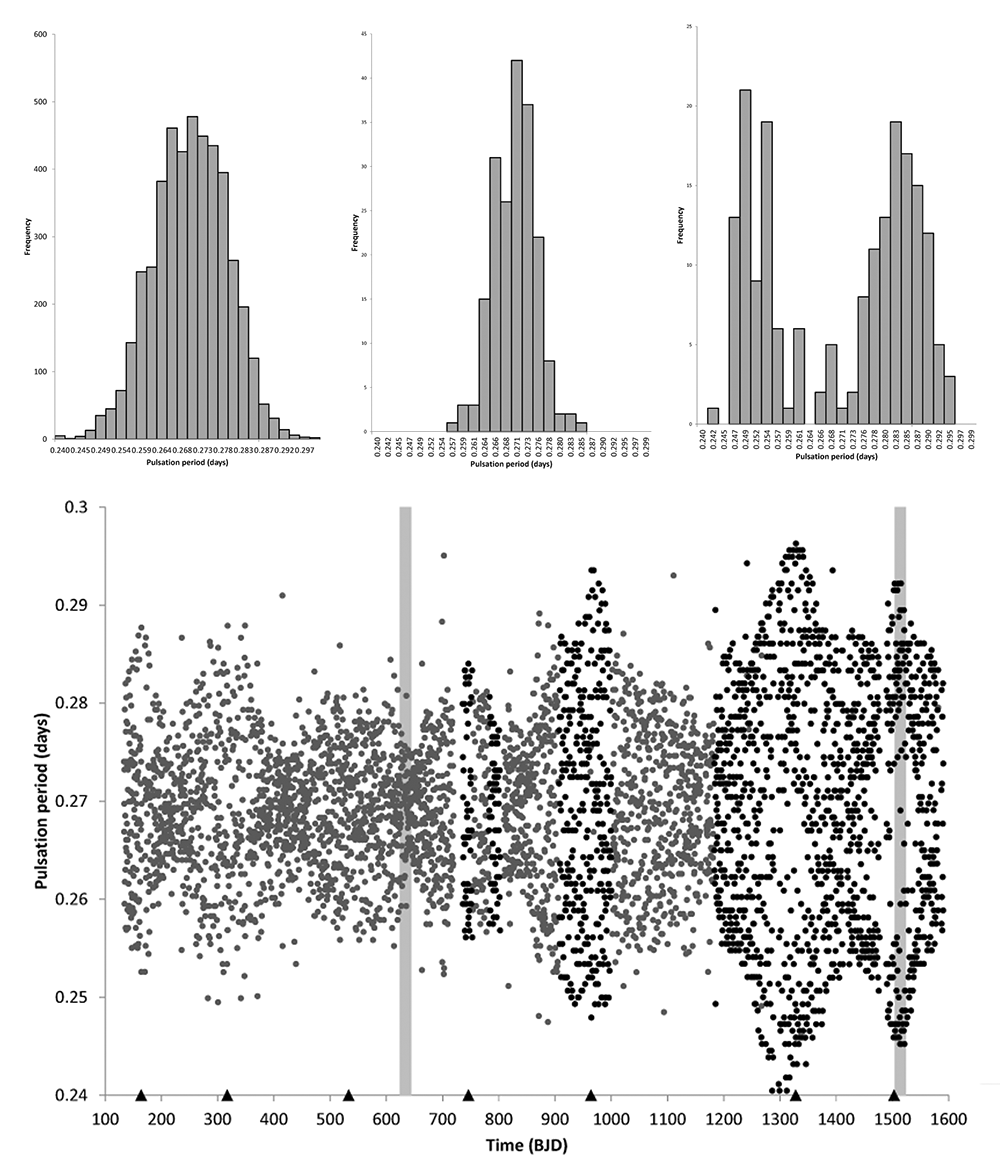}
\end{center}
\caption{\label{fig:fig_7}Histograms of pulsation period length for the complete data set (upper left panel), for BJD=634, a time of low variation (upper middle panel) and for BJD=1514, when variation was high (upper right panel). Another histogram for BJD=1310 can be found in section 5, together with a comparison to simulated data. Times of histograms are indicated with grey bars in the lower panel.}
\end{figure}

The period lengths of the 4,707 cycles that we measured are not just Gaussian distributed, and the distribution changes over time considerably. Fig.~\ref{fig:fig_7} shows their change over time. For every given time, we can see a regime of shorter and a regime of longer period lengths. For the times of higher variance, these regimes are clearly split in two, with no periods of average length (forbidden zone). During times of lower variance, no clear split can be seen.

\clearpage

\subsection{Discussion of period lengths}
It has been proposed that a sufficiently advanced civilization may employ Cepheid variable stars as beacons to transmit all-call information throughout the galaxy and beyond \citep{Learned2008}. The idea is that Cepheids and RR Lyrae are unstable oscillators and tickling them with a neutrino beam at the right time could cause them to trigger early, and hence jog the otherwise very regular phase of their expansion and contraction. The authors’ proposition was to “search for signs of phase modulation (in the regime of short pulse duration) and patterns, which could be indicative of intentional signaling.” Such phase modulation would be reflected in shorter and longer pulsation periods. We showed that the histograms do indeed contain two humps of shorter and longer periods. In case of an artificial cause for this, we would expect some sort of further indication. As a first step, we have applied a statistical autocorrelation test to the sequence of period lengths. This will be discussed below. Furthermore, we have assigned ``one'' to the short interval and ``zero'' to the long interval, producing a binary sequence. This series has been analyzed regarding its properties. It indeed shows some interesting features, including non-randomness and the same autocorrelations. We have, however, found no indication of anything ``intelligent'' in the bitstream, and encourage the reader to have a look himself or herself, if interested. All data are available online (see appendix). A positive finding of a potential message, which we did not find here, could be a well-known sequence such as prime numbers, Fibonacci, or the like. 

\begin{figure}[ht]
\begin{center}
\includegraphics[width=\textwidth]{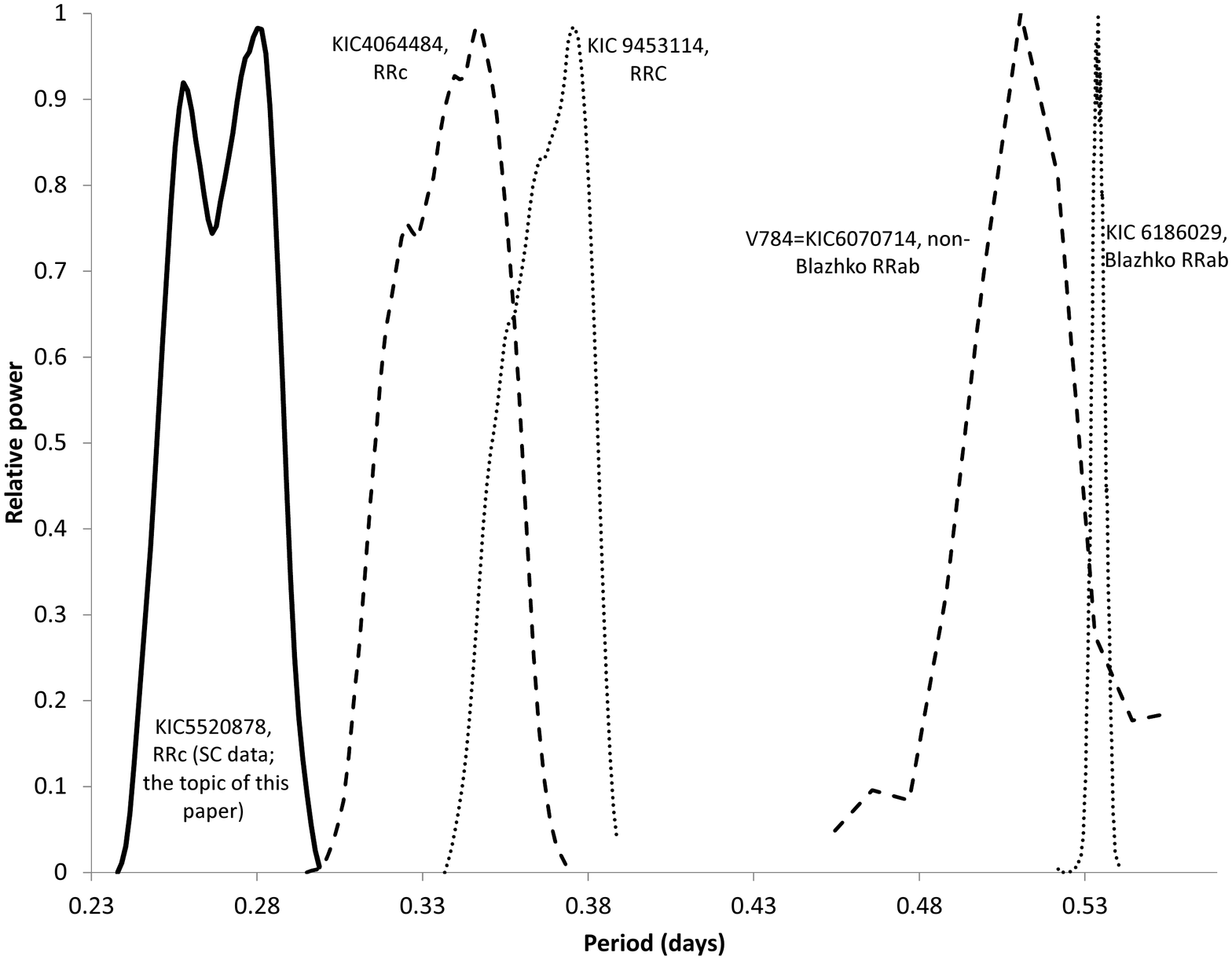}
\end{center}
\caption{\label{fig:f8}Kernel density of pulsation periods for some Kepler RR Lyrae. Only KIC5520878 (on the left) shows a clear double-peak feature.}
\end{figure}

We have also compared several other RR Lyrae stars with Kepler data for their pulsation period distribution, and created Kernel density estimate \citep{Rosenblatt1956,Parzen1962} plots (Fig.~\ref{fig:f8}). We found no second case with two peaks, although some other RRc stars come close.

\section{Prime numbers have high absolute autocorrelation}
\label{sec:primeAC}

\begin{figure}[ht]
\begin{center}
\includegraphics[width=0.6\textwidth]{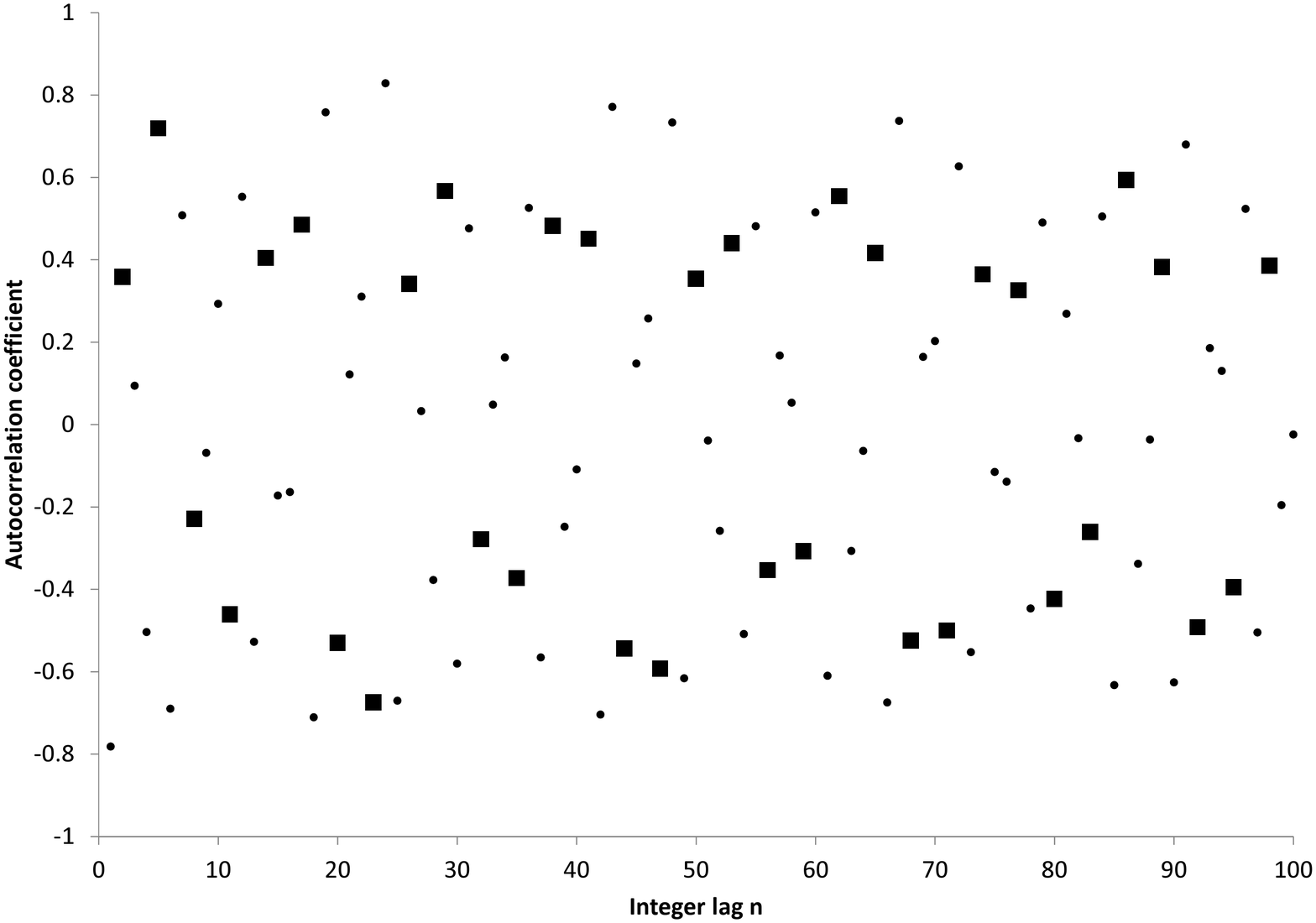}

\includegraphics[width=0.6\textwidth]{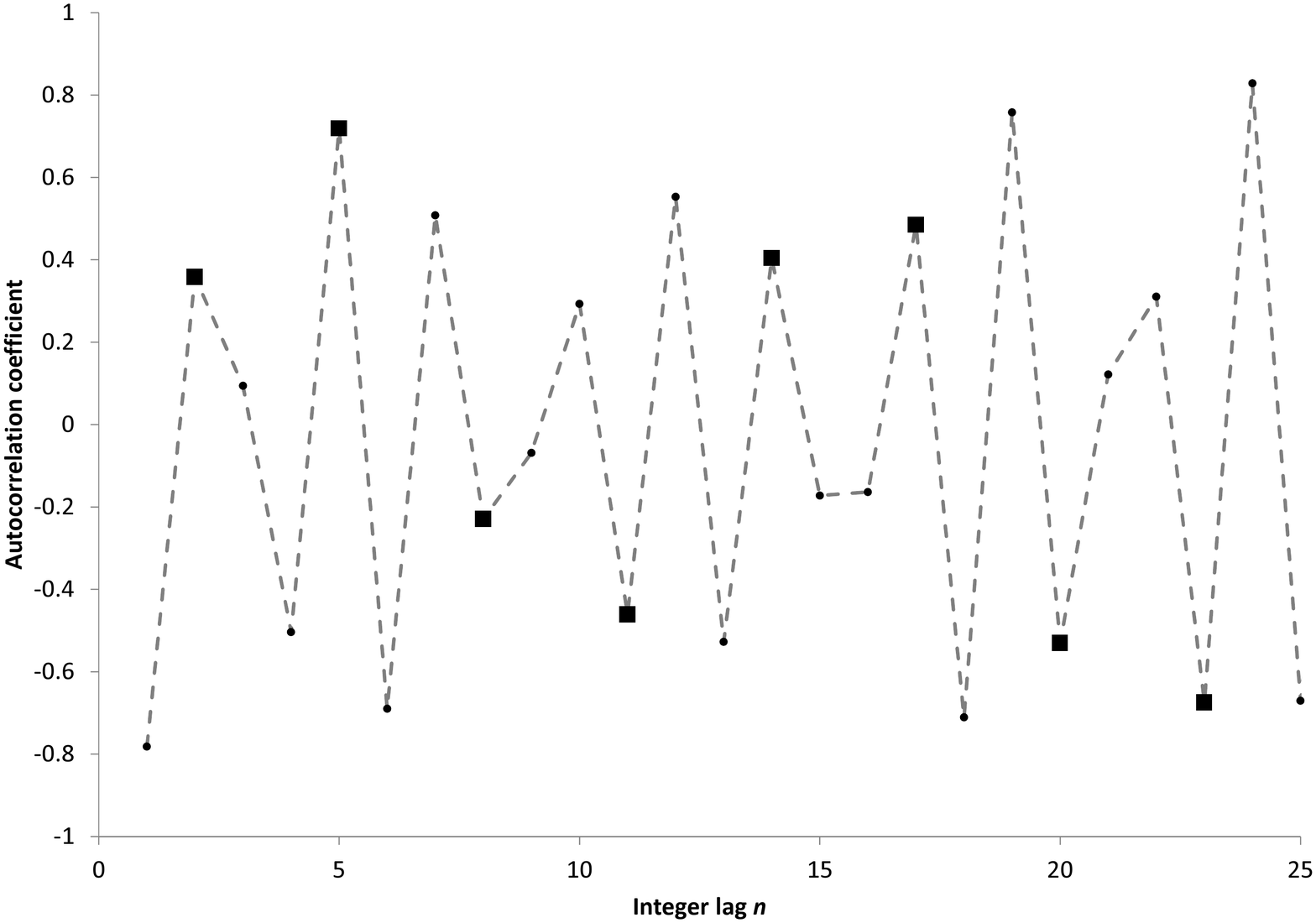}
\end{center}
\caption{\label{fig:f9}Autocorrelation graph for period. All lags $[1..100]$ (top panel). Prime numbers are shown as squares, other numbers as dots. Prime numbers have higher autocorrelation (positive or negative) as other numbers. The average of absolute AC values for the primes is 0.51 ($n$=25), compared to 0.37 ($n$=75) for all others. The difference is significant in a t-test with $p$=99.8\%. The bottom panel shows a zoom into the range of lags $[1..25]$, connecting the integers with dashed lines to show their progression with period-doubling effects.}
\end{figure}

We have performed a statistical autocorrelation analysis for the time series of period lengths. In the statistics of time series analysis, autocorrelation is the degree of similarity between the values of the time series, and a lagged version of itself. It is a statistical tool to find repeating patterns or identifying periodic processes. When computed, the resulting number for every lag $n$ can be in the range [-1, +1]. An autocorrelation of +1 represents perfect positive correlation, so that future values are perfectly repeated, while at -1 the values would be opposite. In our tests, we found that pulsation variations are highly autocorrelated with highest lags for $n$=5, 19, 24, and 42. When plotting integers versus their autocorrelation coefficients (ACFs, Fig.~\ref{fig:f9}), it can clearly be seen that prime numbers tend to have high absolute ACFs. The average absolute ACF of primes in [1..100] is 0.51 (n=25), while for non-primes it is 0.37 (n=75). The average values for these two groups are significantly different in a binominal test (this is the exact test of the statistical significance of deviations from a theoretically expected distribution of observations into two categories), $p=99.79$\% for the SC data, and $p=99.77$\% for the LC data. The effect is present over the whole data set, and for integers up to $\sim$200. As there is no apparent physical explanation for this, we were tempted to suspect some artificial cause. In the following section, however, we will show its natural origin and why primes avoid the ACF range between -0.2 and 0.2.

\clearpage

\subsection{Natural origin of prime numbers}
\label{sec:prime}
\begin{figure}[ht]
\includegraphics[width=0.5\textwidth]{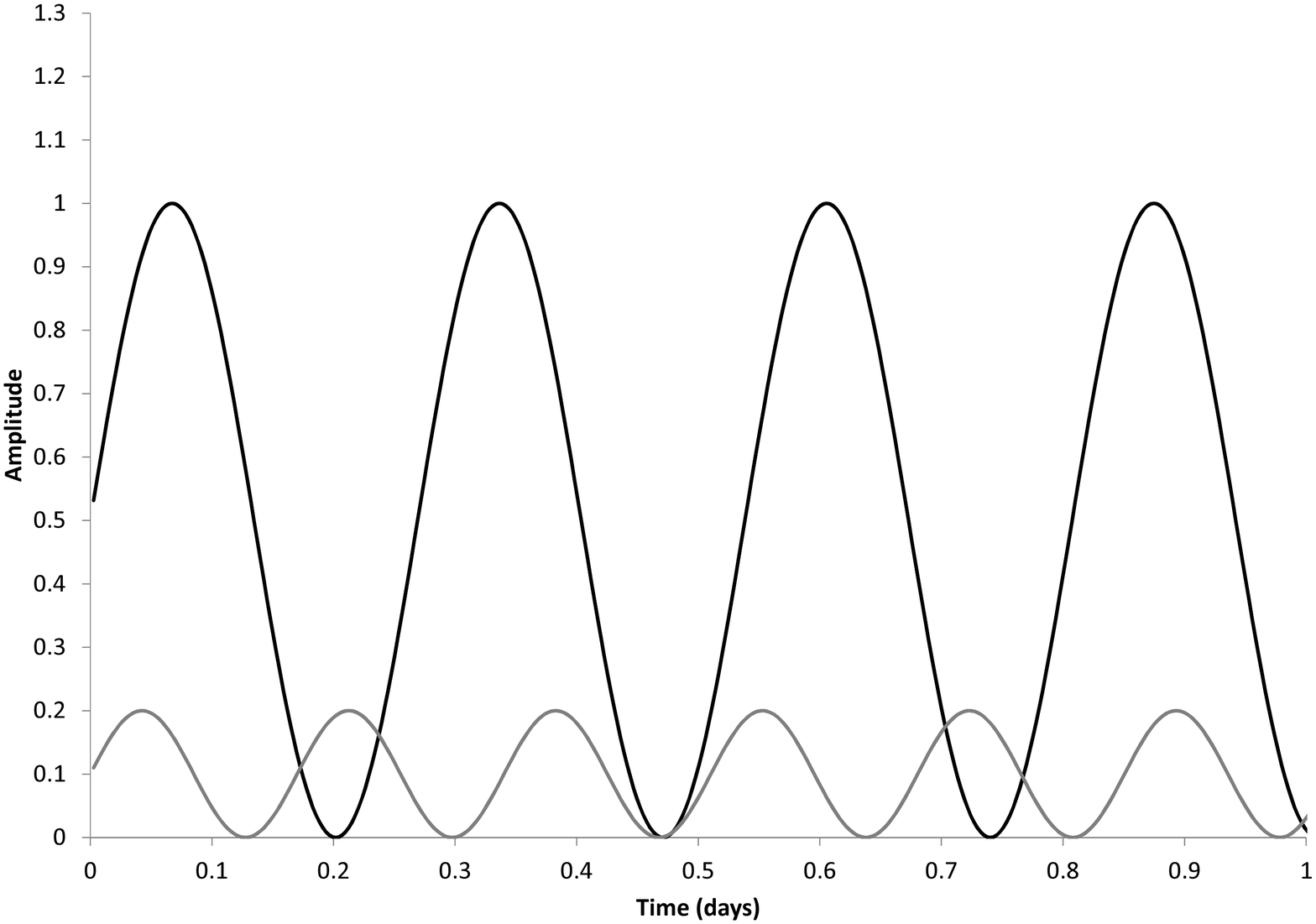}
\includegraphics[width=0.5\textwidth]{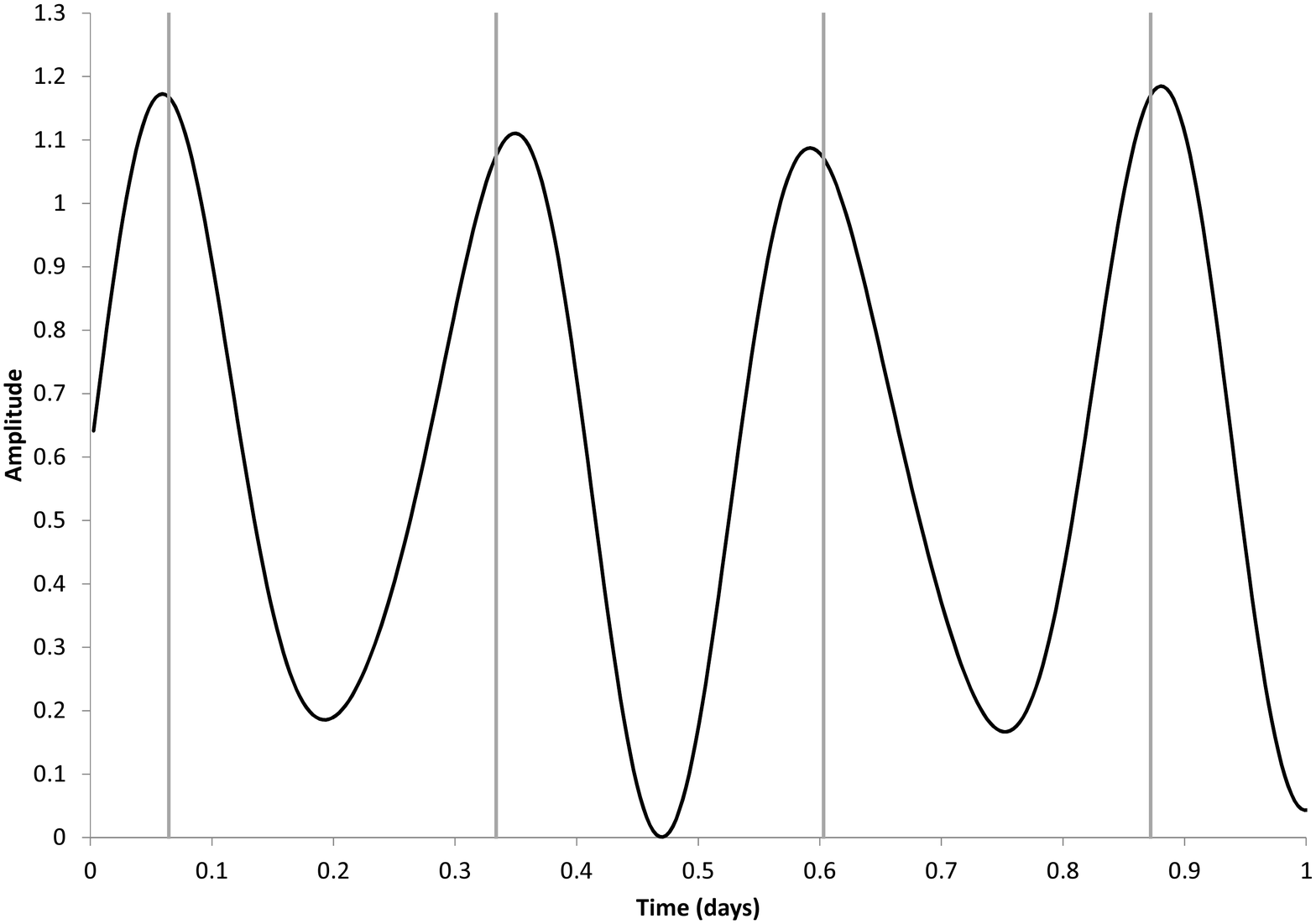}
\caption{\label{fig:f10}Simultaneous sine curve model. Both curves are shown separately in the left panel, and summed up in the right panel.}
\end{figure}

As explained in section 3, through Fourier analysis two pulsation periods can be detected, $P_{1}=0.269$ days and $P_{X}=0.17$ days. While it is clear from figure 1 that the light curve is not a sine curve, for simplicity we have taken it to be the sum of two sine curves in the following model.
We used two sine curves of periods $P_{1}$ and $P_{X}$, with relative strengths of 100\% and 20\%, as shown in Fig.~\ref{fig:f10} (left panel). When added up (right panel), period variations can easily be seen as indicated by the gray vertical lines.

\begin{figure}
\includegraphics[width=\textwidth]{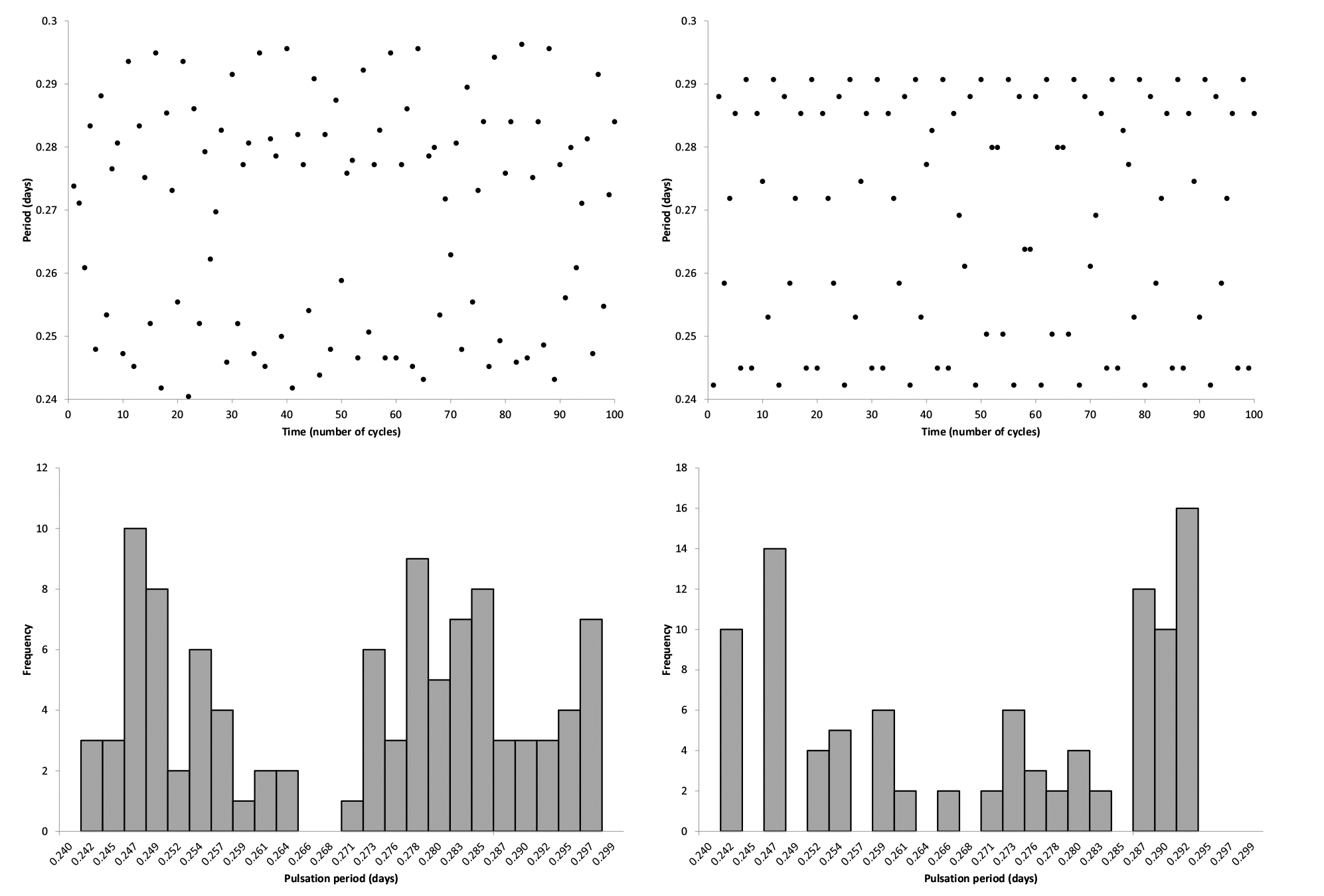}
\caption{\label{fig:f11}Observed (top left) and modeled (top right) period lengths, and their respective histograms (bottom). The observed data is at BJD=1310}
\end{figure}

We then ran this simulated curve for 100 cycles, and compared the peak times with the real data. The comparison is shown in Fig.~\ref{fig:f11} -- this simple model resembles the observational data to some extent. The histogram (figure 13, bottom right) also shows the two humps. When certain ratios of $P_{1}$ and $P_{X}$ are chosen, e.g. $P_{X}/P_{1}=0.632\pm0.001$, it can also be shown that the two simultaneous sines create specific autocorrelation distributions. Let us use a simple model equation of $f(x)=\text{int}(xR)-xR$, where $int$ denotes the rounding to the next integer, and $R$ the period ratio. The function then gives values of close to 0 or close to 0.5 for odd numbers, while for even numbers it gives other values. When re-normalized to the usual ACF range of [-1, +1], this gives high absolute autocorrelation for odd numbers, and less so for even numbers.

We have compared our observational data to this model, and calculated all ACFs(observed) minus ACFs(calculated). The result is an ACF distribution in which primes are still prominent, but not significantly so ($p=91.35$\%). We judge this to be the explanation of the prime mystery. We have also checked the Fourier decomposition for the corresponding amplitudes, and find that $P_{X}$ alone only has an amplitude of 5.1\% compared to $P_{1}$, or 2\% after pre-whitening. However, the $P_{X}$-mode is distributed into multiple peaks, harmonics and linear combinations. In a simultaneous fit to all frequencies, we find a total of 12.7\% for all these peaks that are connected with $P_{X}$ (see table~\ref{tab:fourier}). This is in the required range of the model described above.

\begin{table}
\begin{center}
\caption{Strongest Fourier frequencies\label{tab:fourier}}
\begin{tabular}{lcccc}
\tableline
\tableline
Name\tablenotemark{a} & Frequency & Period & Amplitude\tablenotemark{b} & Amplitude\\
  &   & (days) & (e\textsuperscript{-}/sec) & (percent of $f_{1}$)\\
\tableline
$f_{1}$              & 3.71515  & 0.26917 & 4,990 & 100.0 \\
$2f_{1}$             & 7.43029  & 0.13458 & 549 & 11.0 \\
$3f_{1}$             & 11.14544 & 0.08972 & 308 & 6.2 \\
$f_{X}$              & 5.87884\tablenotemark{c} & 0.17010 & 255 & 5.1 \\
$4f_{1}$             & 14.86058 & 0.06729 & 145 & 2.9 \\
$f_{1}+f_{X}$        & 9.59399  & 0.10423 & 119 & 2.4 \\
$f_{Z}=0.4996 f_{X}$ & 2.93760  & 0.34041 & 83  & 1.7 \\
$2f_{1}+f_{X}$       & 13.30913 & 0.07514 & 37  & 0.7 \\
$f_{X}-f_{1}$        & 2.16370  & 0.46217 & 36  & 0.7 \\
$f_{Z'}=1.499f_{X}$  & 8.81238  & 0.11348 & 27  & 0.5 \\
$f_{1}+f_{Z'}$       & 12.52753 & 0.07982 & 23  & 0.5 \\
$f_{1}-f_{Z}$        & 0.77755  & 1.28609 & 13  & 0.2 \\
$2f_{X}$             & 11.75768 & 0.08505 & 10  & 0.2 \\
$2f_{1}+f_{Z'}$      & 16.24267 & 0.06157 & 10  & 0.2 \\
$f_{1}+f_{Z}$        & 6.65274  & 0.15031 & 10  & 0.2 \\
$f_{1}+2f_{X}$       & 15.47283 & 0.06463 & 7   & 0.1 \\
$2f_{1}+2f_{X}$      & 19.18797 & 0.05212 & 5   & 0.1 \\
\tableline  
All secondary pulsations & & & 634 & 12.7 \\

\end{tabular}

\tablenotetext{a}{Frequencies are taken from \citet{Moskalik2014}}
\tablenotetext{b}{In a simultaneous fit of all frequencies.}
\tablenotetext{c}{We get $P_{X}$/$P_{1}$=0.63195208.}

\end{center}
\end{table}

\clearpage

\section{Visualization of the secondary pulsation}
\begin{figure}
\includegraphics[scale=0.55]{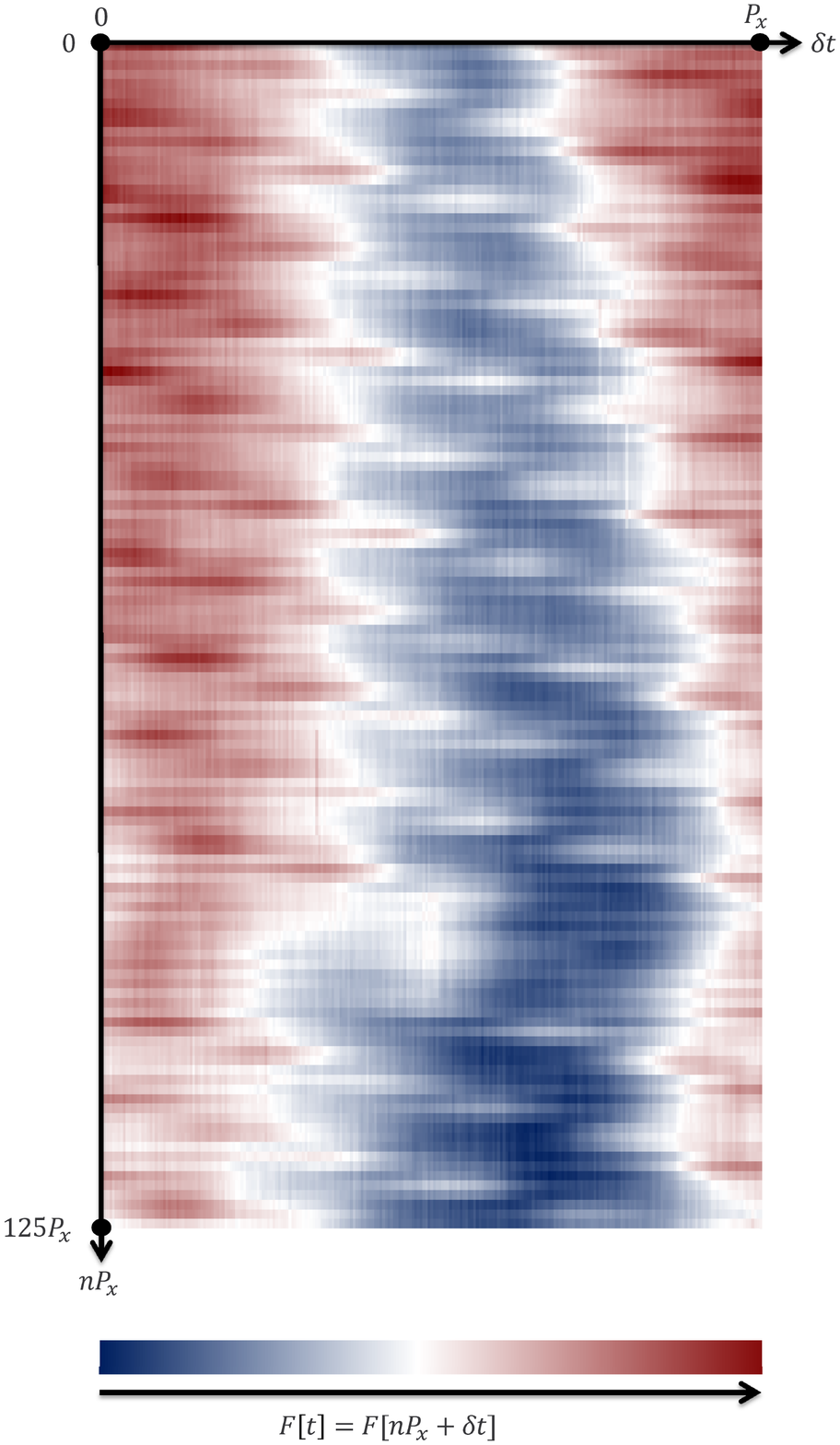}
\caption{\label{fig:heatmap}Heatmap of time-series photometry after subtraction of all frequencies from table~\ref{tab:fourier}, keeping only $P_{X}$. The horizontal axis shows $1P_{X}=0.17d$, the vertical axis is 125 such cycles ($\sim$21.3d) long. Red areas are higher flux. It is clearly visible that $P_{X}$ varies in amplitude and phase. Also, a 9-cycle pattern is apparent.}
\end{figure}

The main conclusion so far is that very strong cycle-to-cycle variations are present, which by far exceed the modulation-induced variations observed in RR Lyrae stars. For certain time periods, a split into a two-peaked distribution can be seen. It is assumed that this originates from a varying $P_{X}$-mode. Autocorrelations can quantify patterns in this, and the O-C diagram (Fig.~\ref{fig:f7}) presents the total effect over time. These tools are useful, but do not show the real mechanism below the surface. To shed more light on the nature of $P_{X}$, we have generated a heatmap (Fig.~\ref{fig:heatmap}). 
For this, we have subtracted all frequencies from table~\ref{tab:fourier}, keeping only $P_{X}$. Due to remaining jitter in the fit, we had to smooth over several cycles to bring out $P_{X}$. The mode shows clear signs of phase and amplitude variation (as expected), and an additional kidney-shaped pattern with a length of nine $P_{X}$-cycles is visible. We consider this heatmap to be the deepest possible visualization of the root cause of the cycle-to-cycle period variations.

\clearpage

\section{Possible planetary companions}
One might raise the question whether some of the observed perturbations are caused by planetary influence, either through transits, or indirectly by gravitational influence of an orbiting planet, or massive companion. With regard to the observed secondary frequency $P_{X}=0.17d$, it seems rather on the short side compared to the almost 2,000 planets discovered so far, but a few of these fall in the same period range \citep{Rappaport2013}. This opens the possibility of perturbations visible as non-radial modes, perhaps caused by gravitationally deforming the star elliptically. Another possible mechanism observed in our own sun concerns the relation between the barycenter for the star and its convection pattern \citep{Fairbridge1987}. 

However, the evolutionary state of RR Lyrae makes it unlikely for them to possess planets. These stars are old and after exhausting their core hydrogen, have gone through the red giant phase and possessed very extended envelopes (up to 20-100 $R_{\sun}$\citep{Smith2004}). Theoretical calculations by \citet{Villaver2009} predicted that no planet of 1M$_{J}$ could survive around a giant 2M$_{\sun}$ star closer than 2.1 AU. Observational data loosens this limit, but the distance distribution of the $\sim$50 planets known around giant stars shows a hard cut-off at 0.5 AU \citep{Jones2014}. Furthermore, the variable luminosity of RR Lyrae must make the evolution of life as we know it on (likely rare) orbiting planets very difficult. During a star's evolution, the habitable zone shifts dramatically, so that any lifeforms would need to accommodate, e.g. by moving to other planets or by moving their habitat. On the other hand, RR Lyrae are older than 10 Gyr, giving potential life more time for evolution than on Earth. A last speculative possibility would be that an advanced civilization would originate from some other nearby more average star and travel to an RR Lyrae for a modulation mission.

\section{Simple Pulsation Models}
We provide two very simple dynamical models to elucidate the pulsations described above, especially with regard to the frequency content and the asymmetry of the light curve. While we hope that these phenomenological models capture some of the essence of the observed pulsation variations, they in no way substitute for well-developed, detailed hydrodynamics models, such as the Warsaw or Florida-Budapest codes \citep{Smolec2008}, which can properly follow the pulsations inside the star and, for example, naturally explain the bumps in the light curve by incorporating shockwaves in the stellar layers. Like the descriptive formalism of Benk\H{o} and colleagues \citep{Benko2011}, our models are intended as simple dynamical analogues, which aim to mathematically reproduce the radial motions of the star while remaining agnostic about the deep underlying physics, including the effects of factors like opacity and temperature variations.

\begin{figure}[ht] 
	\includegraphics[width=0.9\linewidth]{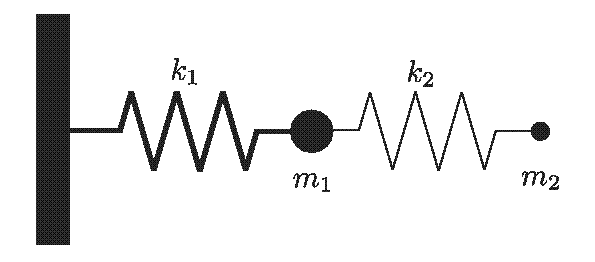} 
	\caption{Lumped masses-versus-springs model of a radial slice of the star with a stiff and dense inner region (left) and a relaxed and rarefied outer region (right).}
	\label{BallSpring}
\end{figure}

\subsection{Masses Versus Springs}	
We first grossly simplify the star into two regions, a stiff dense inner one and a relaxed rarefied outer one. We model these regions with two masses and two springs connected to a fixed wall, as in Fig.~\ref{BallSpring}. For equilibrium coordinates $x_n[t]$, the equations of motion are
\begin{subequations} \label{LumpedEq}
	\begin{align}
		m_1 \ddot x_1 &= -k_1 x_1 - k_2 \left( x_1 - x_2 \right), \\
		m_2 \ddot x_2 &= - k_2 \left( x_2 - x_1 \right).
	\end{align}
\end{subequations}
We fix initial conditions $x_1[0] = 1$, $x_2[0] = 2$, $\dot x_1[0] = 0$, $\dot x_2[0] = 0$ and vary the stiffness and inertia parameters.  For $k_1 = 9$, $m_1 = 8 m$, $k_2 = 1$, and $m_2 = m$, Eq.~(\ref{LumpedEq}) has the solution
\begin{subequations}
	\begin{align}
		x_1 &= +\frac{1}{3} \cos \left[ \omega_{+} t \right] + \frac{2}{3} \cos \left[ \omega_{-} t \right],\\
		x_2 &= -\frac{2}{3} \cos \left[ \omega_{+} t \right] + \frac{8}{3} \cos \left[ \omega_{-} t \right],
	\end{align}
\end{subequations}
where the frequency quotient
\begin{equation}
	\frac{ \omega_{+} }{ \omega_{-} }
	= \frac{ \sqrt{3/2\, m} }{ \sqrt{3/4\, m} }
	=\sqrt{2}
	\approx \frac{1}{0.707}
\end{equation}
is Pythagoras' constant, the first number to be proven irrational. For $k_1 = 4$, $m_1 = 4 m$, $k_2 = 4/5$, and $m_2 = m$, Eq.~(\ref{LumpedEq}) has the solution
\begin{subequations} \label{GoldenRatioEq}
	\begin{align}
		x_1 &=  \frac{5-\sqrt{5}}{10} \cos \left[ \Omega_{+} t \right] + \frac{5+\sqrt{5}}{10} \cos \left[ \Omega_{-} t \right],\\
		x_2 &= \frac{5-3 \sqrt{5}} {5}  \cos \left[ \Omega_{+} t \right] + \frac{ 5+3 \sqrt{5} }{5} \cos \left[ \Omega_{-} t \right],
	\end{align}
\end{subequations}
where the frequency quotient
\begin{equation}
	\frac{ \Omega_{+} }{ \Omega_{-} }
	= \frac{ \sqrt{ \left(\sqrt{5} + 1 \right) / \sqrt{5}\, m } }{ \sqrt{ \left(\sqrt{5} - 1 \right) / \sqrt{5}\, m } }
	= \frac{ 1 + \sqrt{5}  }{2}
	\approx \frac{1}{0.618}
\end{equation}
is the golden ratio, the \textit{most} irrational number (having the slowest continued fraction expansion convergence of any irrational number). In each case, the free mass parameter $m$ allows the individual frequencies to be tuned or scaled to any desired value.

\begin{figure}[ht] 
\includegraphics[width=\textwidth]{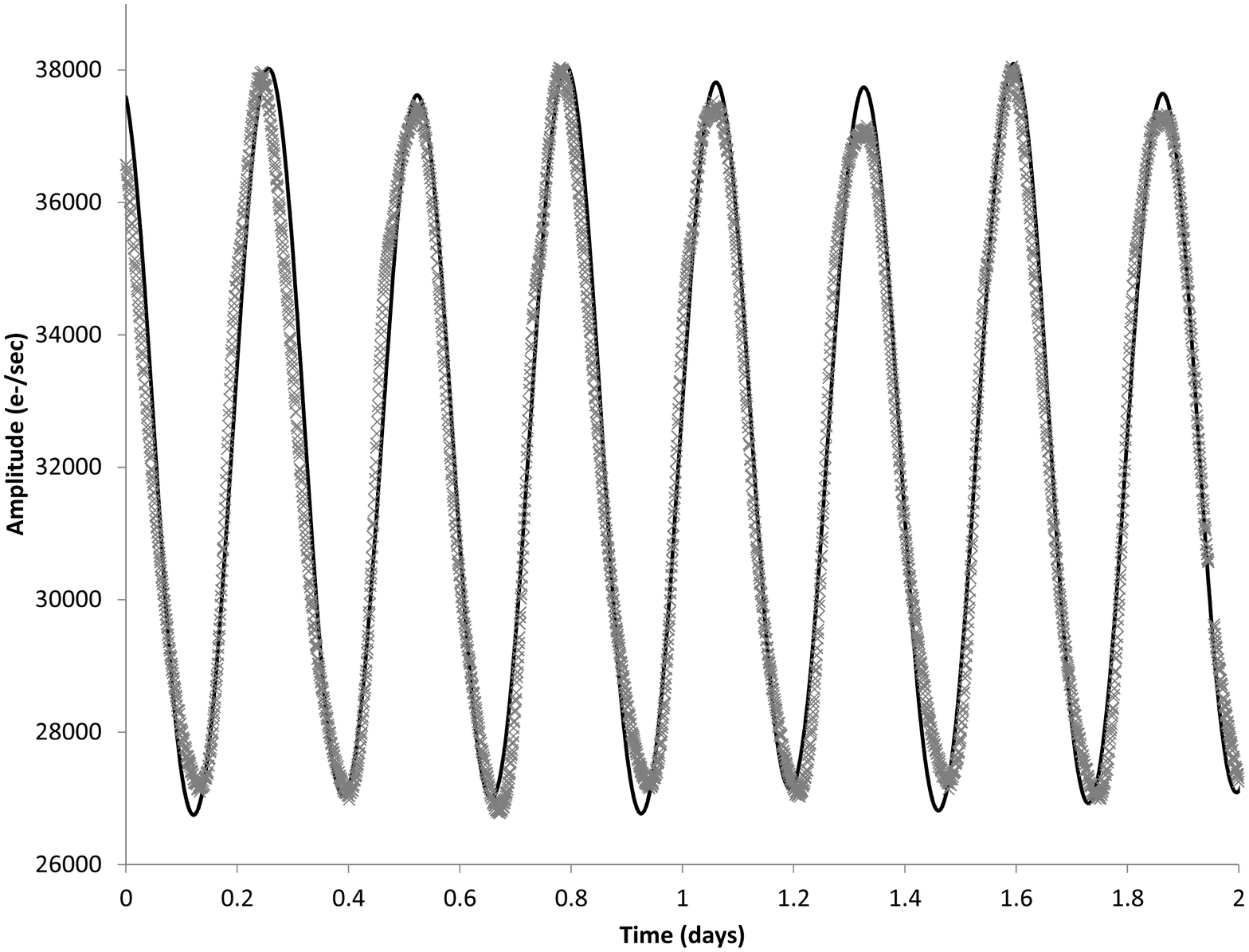}
	\caption{Stellar flux versus time for the lumped masses-versus-springs model (black),  based on the Eq.~(\ref{GoldenRatioEq}) motion $x_2$ of the outer mass with frequencies in the golden ratio, superimposed on Kepler data (grey).}
	\label{OuterSurface}
\end{figure}

We identify the star's radius with the outer mass position, $r = r_e + x_2$, where $r_e$ is the equilibrium radius. If the star's electromagnetic flux is proportional to its surface area, then
\begin{equation} \label{FluxSimpleEq}
	F = k ( r_e + x_2)^2.
\end{equation}
(Squaring generates double, sum and difference frequencies, including those in the original ratios.) For appropriate initial conditions, Fig.~\ref{OuterSurface} graphs the Eqs.~(\ref{GoldenRatioEq}-\ref{FluxSimpleEq}) stellar flux as a function of time and is in good agreement with the Kepler data.

\subsection{Pressure Versus Gravity}	
A star balances pressure outward versus gravity inward. Assuming spherical symmetry for simplicity, the radial force balance on a shell of radius $r$ and mass $m$ surrounding a core of mass $M$ is
\begin{equation} \label{PressureGravityEq}
	m \ddot r = f_r = 4 \pi r^2 P - \frac{G M m}{r^2},
\end{equation}
where  the volume
\begin{equation}
	\frac{V}{V_0} = \left( \frac{r}{r_0} \right)^3,
\end{equation}
and the pressure
\begin{equation}
	\frac{P}{P_0} = \left( \frac{V_0}{V} \right)^\gamma,
\end{equation}
Assuming adiabatic compression and expansion and including only translational degrees of freedom, the index $\gamma = 5/3$, and the radial force
\begin{equation} \label{RadialForceEq}
	f_r =  f_0 \left( \frac{r_e}{r} \right)^2 \left( \frac{r_e}{r} - 1 \right),
\end{equation}
where $r_e$ is the equilibrium radius, so that $f_r[r_e] = 0$, and $f_r[0.68 r_e] \approx f_0$. The corresponding potential energy
\begin{equation}
	U = U_e \left( \frac{r_e}{r} \right) \left( 2 - \frac{r_e}{r} \right),
\end{equation}
where $U_e = -f_0 r_e /2$ is the equilibrium potential energy. The apparent brightness or flux of a star of temperature $T$ at a distance $d$ is proportional to the star's radius squared,
\begin{equation} \label{FluxEq}
	F 
	= \frac{L}{4\pi d^2} 
	= \frac{4 \pi r^2  \left( \sigma T^4 \right) }{4\pi d^2}  
	= \left( \frac{r}{d} \right)^2 \left( \sigma T^4 \right) 
	= k r^2,
\end{equation}
where $k$ is nearly constant if the star's luminosity $L$ depends only weakly on its temperature $T$. (In contrast, \citep{Bryant2014} relates a pulsating star's luminosity to the \textit{velocity} of  its surface.)
  
\begin{figure}[ht] 
\includegraphics[width=0.5\textwidth]{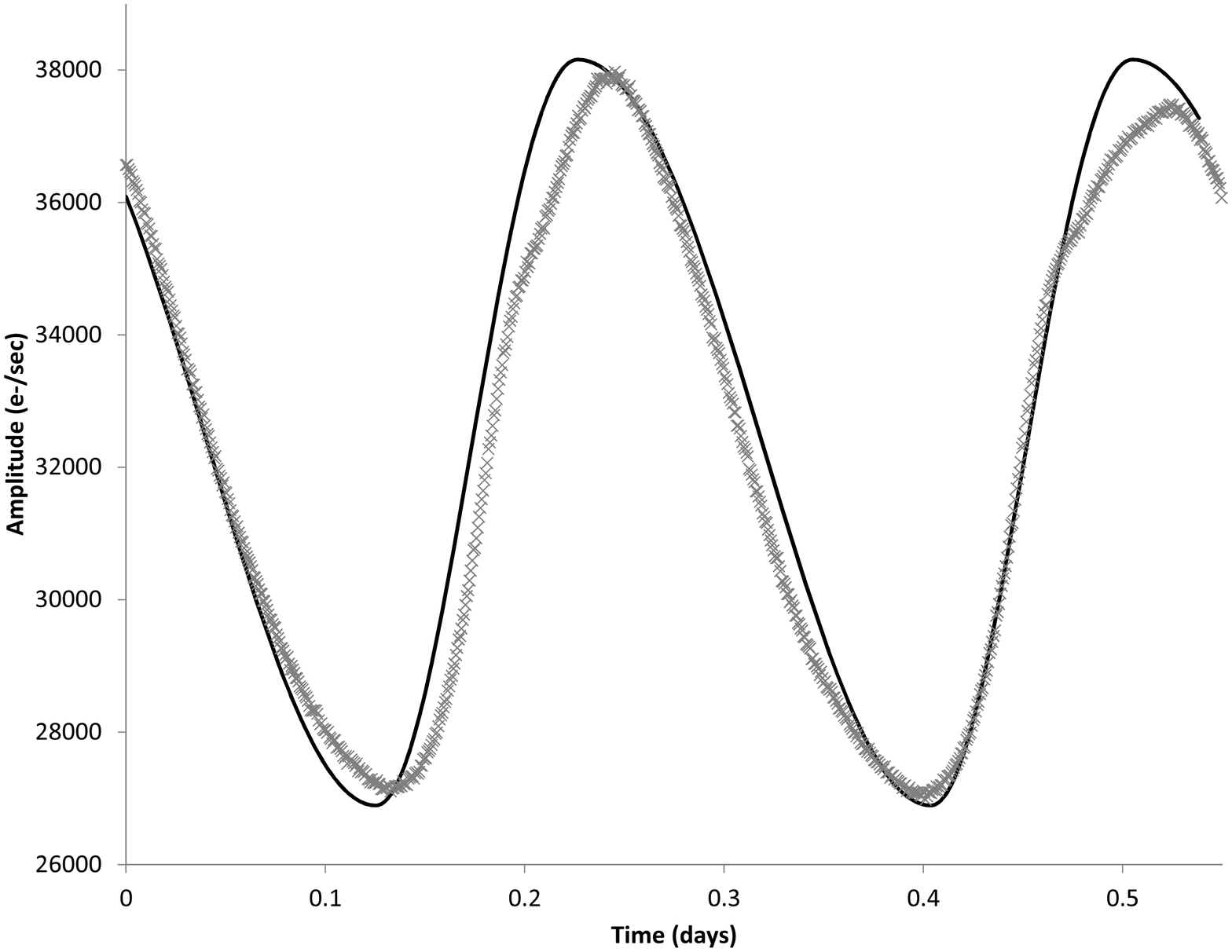}
\includegraphics[width=0.5\textwidth]{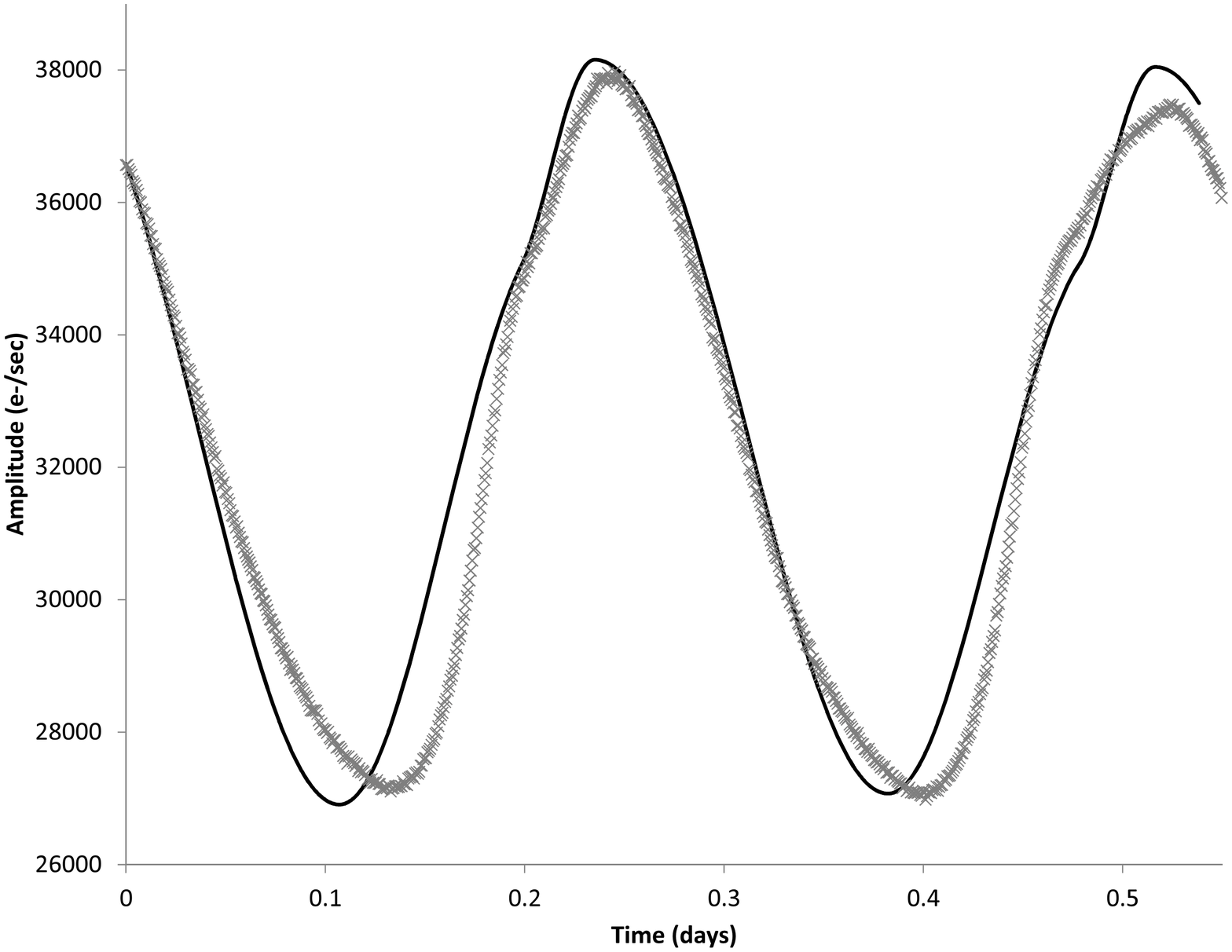}
	\caption{Stellar flux versus time for the pressure-versus-gravity models (black) superimposed on Kepler data (grey). The two-region piecewise constant model asymmetrizes expansion and contraction (left), while the three-region piecewise constant model adds an expansion bump (right).}
	\label{ThreePlots}
\end{figure}

The resulting flux is periodic but non-sinusoidal. The curvature at the maxima is less than the curvature at the minima due to the difference between the repulsive and attractive contributions to the potential, but the contraction and expansion about the extrema are symmetric. To distinguish contraction from expansion, we introduce a velocity-dependent parameter step
\begin{equation} \label{AsymEq1}
	f_0 = \left \{ \begin{array}{ll} 
            	f_1, &  \dot r < 0, \\ 
            	f_2, &\dot r \ge 0.
	\end{array} \right.
\end{equation}
For specific parameters,  the left plot of Fig.~\ref{ThreePlots} graphs the resulting light curve, which asymmetrizes the solution with respect to the extrema. To add a bump to the expansion, we introduce velocity- and position-dependent parameter steps
\begin{equation} \label{AsymEq2}
	p = \left \{ \begin{array}{ll} 
            	p_1, &  \dot r < 0, \\ 
            	p_2, &\dot r \ge 0, r < r_C, \\  
            	p_3, &\dot r \ge 0, r \ge r_C,
	\end{array} \right.
\end{equation}
where $p$ is $f_0$, $r_e$, or $m$. The right plot of Fig.~\ref{ThreePlots} graphs the corresponding light curve. This model captures most of the features of Fig.~5.

From a dynamical perspective, simple models with a minimum of relevant stellar physics are sufficient to reproduce essential features of the KIC 5520878 light curve. The masses-versus-spring example demonstrates that a lumped model can readily be adjusted to exhibit a golden ratio frequency content similar to that of the light curve. The pressure-versus-gravity example demonstrates that an adiabatic ideal gas model can readily be modified to exhibit the asymmetry of the light curve. Nothing exotic is necessary.

\section{Conclusion}
We have tested the idea to search for two pulsation regimes in Cepheids or RR Lyrae, and have found one such candidate \citep{Learned2008}. This fact, together with the large period variations in the range of 20\%, has previously been overlooked due to sparse data sampling. We have shown, however, that the pulsation distribution and the associated high absolute values of prime numbers in autocorrelation in this candidate are likely of natural origin. Our simple modeling of the light curves naturally accounts for their key features.

Despite the SETI null result, we argue that testing other available or future time-series photometry can be done as a by-product to the standard analysis. In case of sufficient sampling, i.e. when individual cycle-length detection is possible, this can be done automatically with the methods described above. Afterwards, the cycle lengths can be tested for a two-peak distribution using a histogram or a kernel density estimate. In the rare positive cases, manual analysis could search for patterns (primes, fibonacci, etc.) in the binary time-series bitstream.

\begin{acknowledgments}
We like to thank Robert Szab\'o and Geza Kovacs for their feedback on an earlier draft, and Manfred Konrad and Gisela Maintz for obtaining the photographs of KIC5520878. We thank the anonymous referee for several useful remarks which improved the quality of this paper. AZ's work is supported by the National Science Foundation under grant no. PHY07-57035 and partially by the Humboldt Foundation. JFL, BK, and WLD gratefully acknowledge support from the Office of Naval Research under Grant No. N000141-21-0026.
\end{acknowledgments}

\appendix*
\section{Appendix: Model Parameters}

Equation~(\ref{GoldenRatioEq}-\ref{FluxSimpleEq}) model parameters and initial conditions for Fig.~\ref{OuterSurface} are

\begin{align} 
	m &= 0.00263,  \nonumber\\
	x_e &= 0.938,  \nonumber\\
	\nonumber\\
	k &= 36\, 562.1, \nonumber\\
	\nonumber\\
	x_{10} &= -0.062, \nonumber\\
	x_{20} &= 0.076, \nonumber\\
	v_{10} &= 0.405, \nonumber\\
	v_{20} &= -0.565. \nonumber
\end{align}

Equations~(\ref{PressureGravityEq}, \ref{RadialForceEq}, \ref{FluxEq}) model parameters and initial conditions for the left plot in Fig.~\ref{ThreePlots}  are

\begin{align} 
	m &= 0.00205,  \nonumber\\
	r_e &= 0.901,  \nonumber\\
	f_1 &= 0.594,  \nonumber\\
	f_2 &= 1.822,  \nonumber\\
	\nonumber\\
	k &= 39\, 160, \nonumber\\
	\nonumber\\
	r_0 &= 0.96, \nonumber\\
	v_0 &= -1. \nonumber
\end{align}
\break
Model parameters and initial conditions for the right plot in Fig.~\ref{ThreePlots} are
\begin{align} 
	m_1 &= 0.00714,  \nonumber\\
	r_{e1} &= 0.904,  \nonumber\\
	f_1 &= -1.36,  \nonumber\\
	\nonumber\\
	m_2 &= 0.005172,  \nonumber\\
	r_{e2} &= 0.8884,  \nonumber\\
	f_2 &= -1.68,  \nonumber\\
	\nonumber\\
	m_3 &= 0.000846,  \nonumber\\
	r_{e3} &= 0.966,  \nonumber\\
	f_3 &= -2,  \nonumber\\
	\nonumber\\
	r_C &= 0.9475, \nonumber\\
	\nonumber\\
	k &= 39\, 040, \nonumber\\
	\nonumber\\
	r_0 &= 0.968, \nonumber\\
	v_0 &= -1.16. \nonumber
\end{align}

\section{Appendix: Pulsation period lengths}
We denote period lengths shorter than the average of 0.269d as 0, otherwise as 1, while missings are marked as X. The first peak is at BJD=133.19818. Data comes from short cadence where available, otherwise long cadence. Line breaks after 48 periods, showing strong positive auto correlation.

%Note to editor: This should be printed in fixed width font (if printed at all)

\begin{lstlisting}
101101010010100101001010101101011010100101011010
101101011010110101011010101101011010100101010010
10100101101010100101011XXXXXXXXXXXXXXXXXX1010110
101101010010XX1011010110101101010110XXXXXXXXX110
10110101010100101001011XX00101010110101010010010
11000101010XX01010010100XX10010111010XX010010100
10XX01100101011XX00001001001XX10100100101XX10110
XXX100XX1001XX10110XX0110XXXXX10XX10011011010010
110100101001011001011001010100101001XX1001XX1010
0XX1001010011101XX00100XXXX1001010XXXXXXXX101011
0101101010101101011XX0110101101010XX110101101011
0101XX111010110XX110XXXX0101XX01101011010XX01011
010101XX1XXXXX01011XX01101010101XX10010100101010
1101010010XX01010010101XX1XX01001010010XX0101010
1XX1XX00101001010XX01010110XXXXXXXXXX101011XX010
10010XX0XX10010101010010XX01010010XX10010101XX1X
X00101001XX00101110101XX1XX101001010XXXXXXXXXXXX
1001010010100010XX01001010010XX01000101001XX0010
1001010XX01110100101XX101001XX100011101011010010
101101100101101011XXXXX0100101000101001011010010
11011010010110101101XXXXXXXXXXXXXXXX101101011010
110010110101101001011010110010110111100101011011
11001110011010010101100101001101011010110110100X
X110110101XX10010010110101101001011010010011XXXX
XX1010110010110100101101001001011010010100101101
0110011010110101101001011XX101101011XX011XX0101X
X101011010100101011010XXXXXX11101011010101001010
010XX11010100101010010100101001010101101XX001010
01010010101011010100XX10001101101XXXXX0101001110
1XX101010010100101001010110101011010100101001010
101101011010100XX1011110101101011010100101010110
10110101101100101XXXXXX0101101010000101001010010
10XX011101011XX001010010100101110100101001010010
10100101001010100101011010XXXXX101101XX001011010
101011010111001101010101101011010110101001010110
101001010101101011XX01001010110XX101101011010110
XXXXX0110XXXX010110101XX101010110101101011010110
011010110101101001001010111010110100101101011000
01101001011010110101XXXX111011100110101100111011
100011000110100101001011100011000110101101101011
100011010110101101011010101011010110100XXX111010
101001010110101101010000101011010110101101010000
101011010110101001010100101011010110101001XX0101
101010XXXXX01010XX010101101010010100101010010101
10101001010010101101010110101001XX0XXX1011010101
0010100101001010101XXXXXXXXXXXXXXXXXXXXXXXXXXXXX
XXXXXXXXXXXXXXXXXXXXXXXXXXX1XX101101010101001010
01010010100101010100101001010010101101010100100X
X010100XX101110010XX010100101XXXXXXXXXXX10100101
001010010101101010100101001010011001101010101101
00XX10010101010101001011010100101011010101101011
010101101011010101101011010110101011000XXXXXXXXX
XXXXXXXXXXXXXX001011010100101001XX01101010110101
01101001010XX010101101011010100101011010XX010101
101010010101101010110101101010010101101010110101
10101XXXX101101010110101101011010101101010110101
101010110101XX101011010XX01010110101001010110101
00101011010101101XXXX101011010110101011010010101
0110101001010110101000010110101011010110101011XX
010010101101011010101001011010101101011XXXX11010
101101011010110101011010100101001010110101011010
100101001010010101011010110101001010010101011010
10010100100XXX1101011010110101001010101101011010
110101101010101101011010110101100110101001011010
11010010011110010101101011010110100XXX1101011010
110111010110100101011010110011010010100101001010
110001010010100101001011101001010010100101001011
101000XXXXX1010010100100010100101001010010101101
01XX001010XX0XXXX0101001010100101001010010101001
01011010101XX1001010101101011XXXX1010XX100101011
010XX0101101010XX11010110101XX101011010100101011
010110XX1011010101101XX1010100100111010101101011
XXXX001010100101011010110XXXXXXXXXXX110101101010
010101001010110101101010010100101010110101001010
01010100101011010110101001010XX110XXXXXX01101XX0
11010101101011010101001010010101101010010XX01010
101101XXXXXXXXX10101101XXXXX0101101010111XXX1010
101101010010101101010110101101010110101001XX0010
101001010110101011010110101011010110101011010110
1XXXXX01011010101101010XXXXXXXXXXXXXXXXXXXX10XXX
1010110100101010100011001010XX01010010101XX10100
101001010XXX010101101011010110101010010101101011
010110101010010101101011010100101010110101101011
010101001010110101101011010101011001XX0101101011
010101001010110101101010110101001010110101101010
101101001010110101101010110101001011100101001010
111XXX011010110101001010110101011010110101001010
1001010110101001010110101011010110100101011XXX10
10110101100XXXX1010110101011010X0010100101010010
101101011010100101010011XXXXXXXXXXXXXXXXXXXXXXXX
XX11010110101001010101101011010111XXXX1101010110
101101011010101011010110101101011010101011010110
101101011010101011010110101101010101101011010110
101100XXXX01101011010110101100010101101011010110
101100010100101001010110101010010100101011010110
1010100101001010110101011010100101001011XXXXXXXX
001010010100101011010101001010010101101010110101
101010010101101010110101001010010101001010110101
0010100101010110100XXX01001010110101011010100101
11XXXXXXXXXXXXXXXXXXXXX1011010101101010010100101
01101010110101001010010101011010110100XXX1101101
010110101001010010101011010110101101010110101101
010110101101010110101011010110101011010101101011
010110101011010100101011011XXXX01011010100101011
00XXXXXXXXXXXXXXXXXXXXXXXXXXXXXXXXXXXXXXXXXX1010
010101101011010100101010110101101011010101101010
110101101011010101101010110101101010110101101010
11010110101011010100101011010101100XXXX101001010
110101011010110101011010101101011010110101011010
101101011010110101010110101101011010010110010110
10110100101000XXXX010111X00101001011010010010110
100101001011010010010110100101101001010010010010
100110XXXXXXXXXXXXXXXXXXX101011010010
\end{lstlisting}


\begin{thebibliography}{}
\bibitem[Alonso et al. (2008)]        {Alonso2008}       Alonso, R., Auvergne, M., Baglin, A. 2008, \aap, 482, L21-L24
\bibitem[Benk\H{o} et al. (2011)]     {Benko2011}        Benk\H{o}, J. M., Szab\'o, R., Paparo, M. 2014,  \mnras, 417, 974
\bibitem[Benk\H{o} et al. (2014a)]    {Benko2014a}       Benk\H{o}, J. M., Plachy, E., Szab\'o, R. 2014, \apjsupp, 213, 2
\bibitem[Benk\H{o} et al. (2014b)]    {Benko2014b}       Benk\H{o}, J. M., Szab\'o, R., Guzik, J. A. et al. 2014, IAU Symposium, 301, 383-384
\bibitem[Blazhko (1907)]              {Blazhko1907}      Blazhko, S. 1907, Astronomische Nachrichten, 175, 327
\bibitem[Bryant (2014)]               {Bryant2014}       Bryant, P. H. 2014, \apjl, 783, L15
\bibitem[Caldwell at al. (2010)]      {Caldwell2010}     Caldwell, D. A., Kolodziejczak, J. J., Van Cleve, J. E. et al. 2010, \apj, 713, L92-L96
\bibitem[Christiansen \& Jenkins (2011)]  {Handbook2011} Christiansen J. L., Jenkins J. M. 2011, KSCI, 19040 - 004
\bibitem[Dworetzki (1982)]            {Dworetzki1982}    Dworetsky, M. M. 1983, \mnras, 203, 917-924
\bibitem[Fairbridge et al. (1987)]    {Fairbridge1987}   Fairbridge, R. W., Shirley, J. H. 1987, Solar Physics, 110, 191-210   
\bibitem[Guggenberger (2013)]         {Guggenberger2012} Guggenberger, E., Kolenberg, K., Nemec, J. M. 2012, \mnras, 424, 649-665  
\bibitem[Jones \& Jenkins (2014)]     {Jones2014} Jones, M. I., Jenkins, J. S., Bluhm, P. et al. 2014, \aap, 566, A113
\bibitem[Kolenberg (2004)]            {Kolenberg2004}    Kolenberg, K. 2004, Proceedings of the International Astronomical Union, 367-372
\bibitem[Kolenberg et al. (2013)]     {Kolenberg2010}    Kolenberg, K., Szab\'o, R., Kurtz, D. W. et al. 2010, \apj, 713, L198  
\bibitem[Koll\'{a}th et al. (2011)]   {Kollath2011}      Koll\'{a}th, Z., Moln\'{a}r, L., Szab\'o, R. 2011, \mnras, 414, 1111-1118
\bibitem[Layden (1995)]               {Layden1995}       Layden, A. C. 1995, \aj, 110, 2288 
\bibitem[Le Borgne et al. (2014)]     {Borgne2014}       Le Borgne, J. F., Poretti, E., Klotz, A. et al. 2014, \mnras, 441, 1435-1443
\bibitem[Learned et al. (2008)]       {Learned2008}      Learned, J. G., et al. 2008, arXiv:0809.0339
\bibitem[Lee (1992)]                  {Lee1992}          Lee, Y.-W. 1992, ed. Warner, B., RR Lyrae Variables and Galaxies Variable Stars and Galaxies, 30, 103
\bibitem[Lomb (1976)]                 {Lomb1976}         Lomb, N. 1976, Kluwer Academic Publishers, 39, 447-462
\bibitem[Marconi et al. (2005)]       {Marconi2005}      Marconi, M., Nordgren, T., Bono, G. et al. 2005, \apj, 623, L133-L13
\bibitem[Moskalik (2014)]             {Moskalik2014}     Moskalik, P. 2014, Precision Asteroseismology, 301
\bibitem[Nemec et al. (2013)]         {Nemec2013}        Nemec, J. M., Cohen, J. G., Ripepi, V. et al. 2013, \apj,  773, 181
\bibitem[Parzen (1962)]               {Parzen1962}       Parzen, E. 1962, The Annals of Mathematical Statistics 33, 3
\bibitem[Petrie (1962)]               {Petrie1962}       Petrie, R. M. 1962, Astronomical Techniques, University of Chicago Press
\bibitem[Pickering and Leavitt (1912)]{Pickering1912}    Leavitt, H. S., Pickering, E. C. 1912, Harvard College Observatory Circular, 173, 1-3
\bibitem[Ransom et al. (2010)]        {Ransom2010}       Ransom, S. M., Eikenberry, S. S., Middleditch, J. 2002, \aj, 124, 1788-1809
\bibitem[Rappaport et al. (2013)]     {Rappaport2013}    Rappaport, S., Sanchis-Ojeda, R., Rogers, L. A. 2013, \apjl, 773, L15  
\bibitem[Rosenblatt (1956)]           {Rosenblatt1956}   Rosenblatt, M. 1956, The Annals of Mathematical Statistics 27, 3
\bibitem[Scargle (1982)]              {Scargle1982}      Scargle, J. D. 1982, \apj, 263, 835-853
\bibitem[Smith (2004)]                {Smith2004}        Smith, H. 2004, RR Lyrae Stars, Cambridge University Press
\bibitem[Smolec and Moskalik (2008)]  {Smolec2008}       Smolec, R., Moskalik, P. 2008, \actaa, 58, 193 
\bibitem[Smolec et al. (2012)]        {Smolec2012}       Smolec, R., Soszynski, I., Moskalik, P. 2012, \mnras, 419, 2407-2423
\bibitem[Stellingwerf et al. (2013)]  {Stellingwerf2013} Stellingwerf, R. F., Nemec, J. M., Moskalik, P. 2013, arXiv:1310.0543
\bibitem[Szab\'o et al. (2010)]       {Szabo2010}        Szab\'o, R., Koll\'{a}th, Z., Moln\'{a}r, L. et al. 2010, \mnras, 409, 1244-1252
\bibitem[Szab\'o et. al. (2011)]      {Szabo2011}        Szab\'o, R., Szabados, L., Ngeow, C.-C. et al. 2011, \mnras, 413, 2709-2720
\bibitem[Szab\'o et. al. (2014)]      {Szabo2014}        Szab\'o, R., Benk\H{o}, J. M., Paparo, M. et al. 2014, \aap, in print, arXiv:0809.0339
\bibitem[Villaver \& Livio (2009)]    {Villaver2009}     Villaver, E., Livio, M. 2009, ApJ, 705, 81
\bibitem[Welch (2014)]                {Welch2014}        Welch, D., Website, \url{http://www.aavso.org/now-less-mysterious-blazhko-effect-rr-lyrae-variables}, accessed 21-Aug-2014
\end{thebibliography}
\end{document}